\documentclass[10pt,leqno]{article}
\usepackage{graphicx}
\usepackage{amsmath, amssymb, amsfonts}
\usepackage{float}
\usepackage{indentfirst,csquotes}
\usepackage[
    a4paper,
    left=0.5in,
    right=0.5in,
    top=0.5in,
    bottom=1.5in
]{geometry}
\usepackage{amssymb,amsthm,amsmath}
\usepackage{xcolor,paralist,hyperref,titlesec,fancyhdr,etoolbox}

\hypersetup{ colorlinks=true, linkcolor=black, filecolor=black, urlcolor=black }
\usepackage{lipsum}
\begin{document}

\title{WKB Analysis of the Astrophysical S-Factor for the pp-Fusion Reaction}
 
\author{
  Arushi Sharma, Ishwar Kant and O.\,S.\,K.\,S.\ Sastri\thanks{Corresponding author: \texttt{sastri.osks@hpcu.ac.in}} \\[4pt]
  \small Department of Physics and Astronomical Sciences, \\
  \small Central University of Himachal Pradesh, Dharamshala, India
}
\date{}
\maketitle


\begin{abstract}
The current S-factor calculations for weak pp interaction involve the determination of low-energy penetration probability using the bare Coulomb Gamow factor, which renders the nature and shape of actual interaction to be insignificant. In this work, the astrophysical S-factor is obtained utilizing the WKB action integral evaluated over the inverse scattering potential for the S-wave of pp-interaction, which does not involve bare Coulomb interaction in it. The np and pp inverse potentials are constructed using the phase function method by providing a reference potential consisting of three smoothly combined Morse functions, whose model parameters are optimized using a genetic algorithm to minimize the mean-squared error between the obtained and expected scattering phase shifts. The overlap integral between the bound-state deuteron and the scattering state of pp S-wave has been evaluated at different energies all the way up to 0.0001 MeV, and corresponding fusion cross-sections are determined. The WKB action integral has been computed at all energies without any approximations. Finally, the S-factor at various energies are calculated, and S(0) has been obtained using a supervised neural network. The value of S(0) obtained using our methodology involving complete evaluation of WKB action integral without approximations is $(0.1261\pm 0.0043)\times10^{-25}$, which is almost one order of magnitude lower than the currently accepted values using various methods. The inverse potentials constructed using the reference potential approach, which does not involve explicit consideration of nuclear $\&$ Coulomb interaction, resulting in a finite range of pp-interaction, have been successful in providing a new estimate for the astrophysical s-factor that depends on the nature and shape of the actual potentials.  
\end{abstract} 

\section{Introduction:-}
The weak fusion reaction

\[
p + p \rightarrow d + e^{+} + \nu_{e}
\]

initiates the proton–proton chain that drives hydrogen burning in main-sequence stars and thus plays a fundamental role in stellar nucleosynthesis, solar energy generation, and neutrino production \cite{Bethe1939,Bahcall2005,Rolfs1988,Iliadis2007}. Since the thermal energies prevailing within stellar interiors are considerably lower than the Coulomb barrier associated with proton–proton fusion, the reaction proceeds primarily through quantum tunneling. Consequently, the corresponding fusion cross section is extremely small in the astrophysical energy regime, rendering direct experimental measurements under terrestrial conditions exceedingly challenging.

At low energies, strong variation of proton–proton fusion cross section is governed predominantly by the penetrability of the Coulomb barrier, while the nuclear interaction itself is conventionally encapsulated within the astrophysical S-factor \cite{Iliadis2007,Gurney1928}. The S-factor separates the comparatively smooth nuclear contribution from the dominant energy dependence arising from Coulomb suppression, thereby providing a more suitable representation of the intrinsic reaction dynamics at astrophysical energies. Accordingly, zero-energy astrophysical S-factor, S(0), constitutes a fundamental quantity in evaluation of thermonuclear reaction rates and the theoretical description of solar energy generation.
\\

The first quantitative calculation of the proton--proton fusion S-factor was carried out by Bethe and Critchfield within the framework of Fermi’s theory of weak interactions, leading to an estimated value of \cite{Bethe1938}
\[
S(0)\approx3.6\times10^{-25}\,\mathrm{MeV\,b}.
\]
Their treatment employed plane-wave proton states together with a simplified deuteron wavefunction and neglected detailed strong-interaction and Coulomb distortion effects. Subsequent investigations by Salpeter improved the treatment of low-energy barrier penetration using Gamow theory, leading to refined estimates of the fusion probability at stellar energies \cite{Salpeter1952}.

Further developments were carried out by Bahcall and collaborators using improved descriptions of the proton--proton scattering state and deuteron bound-state wavefunction based on phenomenological nucleon--nucleon interactions \cite{Bahcall1966,Kamionkowski1994}. Their calculations yielded values in the range
\[
S(0)\approx(3.9\text{--}4.0)\times10^{-25}\,\mathrm{MeV\,b}
\]
and substantially reduced the theoretical uncertainties present in earlier calculations.

Later studies by Schiavilla and collaborators incorporated meson-exchange current contributions in the weak interaction operator and employed realistic nucleon--nucleon interactions in the calculation of the transition matrix element \cite{Schiavilla1998}. These calculations reported values close to
\[
S(0)\approx4.0\times10^{-25}\,\mathrm{MeV\,b}
\]
and demonstrated the importance of two-body current effects in precision calculations of proton-proton fusion.

The development of effective field theory (EFT) provided an alternative framework for calculating the proton--proton fusion S-factor at very low energies \cite{Butler2001,Park2003,Marcucci2013,Liu2022}. Early EFT calculations by Kong and Ravndal reported values of the astrophysical S-factor close to \cite{Kong2001,Kong1999}
\[
S(0)\approx4.0\times10^{-25}\,\mathrm{MeV\,b}.
\]
Later hybrid EFT calculations employing Argonne $v_{18}$ wavefunctions together with weak current operators yielded values around \cite{Park2003,Schiavilla1998,Adelberger2011}
\[
S(0)\approx3.94\times10^{-25}\,\mathrm{MeV\,b}
\]
with significantly reduced theoretical uncertainties. However, these calculations still relied on fitted low-energy constants and phenomenological treatments of certain short-range contributions.

A major consolidation of modern proton--proton fusion calculations was provided by the Solar Fusion II evaluation of Adelberger \textit{et al.} \cite{Adelberger2011}, which critically reviewed the existing potential-model and EFT-based approaches and established a recommended value of
\[
S(0)=(4.01\pm0.04)\times10^{-25}\,\mathrm{MeV\,b}.
\]
This evaluation incorporated realistic nucleon--nucleon interactions, Coulomb distortion effects, meson-exchange currents, radiative corrections, and weak current contributions, and has since served as one of the standard references for solar-model calculations and low-energy weak fusion processes.

More recent calculations based on chiral effective field theory ($\chi$EFT) by Marcucci  and collaborators reported values close to \cite{Marcucci2013}
\[
S(0)\approx4.03\times10^{-25}\,\mathrm{MeV\,b}
\]
with estimated theoretical uncertainties of less than $1\%$. These calculations included higher-order electromagnetic and weak-interaction corrections, although the results still depend on regulator choices and truncation schemes within the EFT expansion.

In 2019, Acharya, Platter, and Rupak investigated higher partial-wave contributions, particularly P-wave effects in proton-proton fusion \cite{Acharya2019}. Their analysis showed that these contributions remain negligible ($<0.5\%$) at solar energies, thereby confirming the dominance of S-wave capture assumed in earlier calculations. The resulting value
\[
S(0)\approx4.047\times10^{-25}\,\mathrm{MeV\,b}
\]
represented only a minor refinement over previous high-precision determinations.
Hybrid EFT  calculations employing the AV18 interaction together with two-body axial current operators constrained from tritium $\beta$-decay reported values close to \cite{Schiavilla1998,Park2003,Walecka1995}
\[
S(0)\approx3.94\times10^{-25}\,\mathrm{MeV\,b},
\]
whereas later chiral EFT calculations yielded values in the range
\[
S(0)\approx(4.0\text{-}4.1)\times10^{-25}\,\mathrm{MeV\,b}
\]
Since direct experimental measurements of the proton--proton fusion cross section at stellar energies are not presently feasible, the quoted uncertainties in these calculations arise primarily from theoretical sources such as EFT truncation effects, regulator dependence, and experimental uncertainties \cite{Adelberger2011,Schiavilla1998,Kamionkowski1994}. 

In most of the modern high-precision calculations discussed above, realistic nucleon-nucleon interactions such as Argonne $v_{18}$ are predominantly employed for description of both proton-proton scattering and deuteron bound states. In these approaches, electromagnetic interaction between the two protons is generally incorporated through the bare Coulomb potential \cite{Kong1999,Joachain1975}
\[
V_C(r)=\frac{e^2}{r},
\]
which is long-range in nature.
Under this approximation, Gamow factor employing WKB integral $ (\exp[-\frac{2}{\hbar}\int_{r_1}^{r_2}\sqrt{2\mu(V(r)-E)}\,dr])$ reduces to the
\[
G(E)\sim e^{-2\pi\eta},
\]
where
\[
\eta=\frac{\mu e^2}{\hbar^2 k}
\]
is the Sommerfeld parameter \cite{Jackson1950} associated with low-energy Coulomb scattering, and $k$ is the relative wave number of the interacting protons. 

This bare coulomb approximation is incorporated through normalization factor in the pp scattering\cite{bethe1949,Jackson1950}.
\[
C_0^2(\eta)=\frac{2\pi\eta}{e^{2\pi\eta}-1},
\]

The corresponding proton-proton scattering wavefunction is then constructed by solving the radial Schr\"odinger equation in the presence of both nuclear and Coulomb interactions and subsequently matching the numerical solution with the asymptotic Coulomb wavefunction at large distances in order to extract the low-energy scattering behavior and phase shifts \cite{Schiavilla1998}. The asymptotic scattering wavefunction is expressed in terms of the regular and irregular Coulomb functions together with the scattering phase shift determined from the asymptotic matching procedure \cite{Joachain1975,Thompson1988}. Similarly, the deuteron bound-state wavefunction is calculated using realistic coupled-channel nucleon--nucleon interactions together with the experimental deuteron binding energy and low-energy deuteron properties \cite{Schiavilla1998,Kamionkowski1994}. In these calculations, the coupled ${}^{3}S_{1}$ and ${}^{3}D_{1}$ channels are solved self-consistently using the experimental deuteron binding energy as an input in order to obtain the corresponding deuteron radial wavefunctions and asymptotic normalization behavior.

Although these calculations achieve remarkable agreement and high numerical precision by employing realistic nucleon–nucleon interaction potentials for the proton–proton scattering and deuteron bound-state wavefunctions, the corresponding barrier penetration probability is still evaluated using the conventional Gamow treatment based on the unscreened bare Coulomb potential $\frac{e^2}{r}$ instead of using total interaction potential in WKB integral.
 At extremely low astrophysical energies, where the fusion probability is strongly governed by barrier-penetration effects, this difference in the underlying interactions between the wavefunction and fusion probability influences the resulting overlap $\Lambda$ and the corresponding S-factor. Also, in the analytic expression for $\Lambda^2$ obtained within the effective-range approximation remains insensitive to the detailed shape of the total effective interaction potential, since the complete interaction potential is not explicitly incorporated in the formalism and the expression depends only on the low-energy scattering parameters \cite{kamionkowski1993rate}.

Since the bare Coulomb interaction does not incorporate screening or effective short-range modifications arising from the effective proton-proton interaction, the resulting Gamow factor may become overestimated at very low energies. However, due to the long-range nature of the Coulomb interaction, the potential does not vanish asymptotically. Consequently, the direct evaluation of the complete WKB penetration integral using the interaction potential becomes increasingly challenging at extremely low astrophysical energies, and it also requires integration over very large distances (up to 3000 fm), making the computation highly intensive. Mostly, conventional calculations retain only the pure Coulomb term in the penetration integral in order to simplify the treatment of the low-energy scattering problem and the corresponding overlap matrix element. In these calculations, the classical turning points and the effective Coulomb barrier used in the calculation of the tunneling probability does not fully represent the complete interaction governing the proton-proton fusion process.

Since the Coulomb potential decreases as 1/r, its contribution remains significant even at large separations and therefore does not vanish asymptotically. As discussed by Taylor \cite{Taylor1974}, the long-range behavior of the Coulomb interaction causes the corresponding phase shifts contribution to continue accumulating over extended radial distances. Consequently, the proton--proton wavefunction obtained by solving the Schr"odinger equation and matched to the Coulomb wavefunctions yields physically reliable low-energy scattering phase shifts. However, when the same interaction potential is used in the phase equation, the long-range Coulomb interaction causes the phase to accumulate over a large radial interval, leading to numerical sensitivity and inaccurate phase shifts at extremely low energies. In particular, for AV18-based proton-proton scattering calculations, stable numerical convergence requires matching distances extending to several thousand femtometers at such low energy.
Furthermore, since approximations based solely on the bare Coulomb potential become increasingly difficult to apply accurately at extremely low energies due to the exponential growth of the inverse Gamow factor $e^{2\pi\eta}$ and the dominance of the $ 1/sqrt(E)$ dependence, the astrophysical S-factor is commonly extrapolated to the zero-energy limit, S(0), using polynomial fitting procedures. \cite{Bahcall1969}. These extrapolations introduce additional approximations in the low-energy analysis, since they rely primarily on low-energy experimental data, which are not available down to energies of $10^{-4}$ MeV. At the same time, uncertainties associated with the underlying effective interaction and potential-dependent tunneling behavior are generally not treated explicitly in most existing calculations.

We have implemented an alternative approach to construct the underlying interaction potential for np\cite{awasthi2024high} and pp\cite{sharma2025novel} scattering using a combination of phase function method, reference potential approach and genetic algorithm-based optimisation \cite{sharma2026genetic, awasthi2025genetic}.

The advantage of the reference potential approach is that it is purely phenomenological and does not explicitly include any nuclear or Coulomb interactions. Hence, one obtain the inverse potential from a family of curves that best fits the expected scattering phase shift. Using a combination of three Morse functions which are smoothly connected at the two boundaries, we have obtained the inverse potential for S-wave pp-scattering that has similar potential depth and Coulomb barrier as that of AV18 but with the potential dying down to zero by about 34 fm. We have previously determined the inverse potential for S-wave of $\alpha-\alpha$ scattering and observed that the potential goes down to zero by about 36 fm \cite{sastri2024constructing}. NLEFT lattice calculations for the same have also shown similar results \cite{elhatisari2015ab}. This inverse potential approach incorporates the screened Coulomb interaction without the need for including an explicit Coulomb function, is the motivation to re-examine the proton-proton fusion tunneling probability and overlap integral using the full interaction within the WKB framework.
In the present work, the proton--proton fusion problem is re-examined by incorporating proper screening effects through the Reference Potential Approach together with a direct evaluation of the WKB tunneling probability using the complete effective interaction potential. 

Instead of employing only a pure Coulomb approximation in the penetration integral, the present formulation utilizes the complete effective proton-proton interaction potential for the determination of the barrier penetration probability, classical turning points, continuum scattering states, overlap integrals, and nuclear transition quantities within a unified effective interaction framework in the extremely low-energy region relevant to stellar proton-proton fusion. Also, the deuteron bound-state wavefunction is calculated by solving the coupled $^3S_1 - ^3D_1$
differential equations using realistic nucleon-nucleon interactions. In addition, uncertainties associated with the deuteron binding energy arising from the underlying data analysis are also investigated. The proton-proton scattering wavefunction is obtained using the Phase Function Method (PFM) \cite{Babikov1967} within a physics-informed optimization framework, where the scattering phase shifts are employed in the determination of the effective proton--proton interaction potential. The corresponding zero-energy extrapolation of the S-factor is performed using supervised neural networks, which provide a more reliable and data-driven alternative to conventional polynomial fitting methods.

\section{Methodology}

The spin-averaged total cross section for the
\({}^{1}\mathrm{H}(p,e^{+}\nu_e){}^{2}\mathrm{H}\)
reaction is expressed as

\begin{equation}
\sigma(E)
=
\frac{1}{(2\pi)^3}
\,
\frac{G_V^2}{v_{\mathrm{rel,n.r.}}}
\,
m_e^5
\,
f_{pp}(E)
\,
\sum_{M}
\left|
\left\langle
d,M
\left|
\mathbf{A}_{\mu}^{(1)}
\right|
pp
\right\rangle
\right|^2 
\label{pcs}
\end{equation}
where 

\[
G_V
=
(1.14939 \pm 0.00065)\times10^{-5}\ \mathrm{GeV}^{-2},
\]
 is the vector coupling constant, \(m_e\) is the electron mass, \(v_{\mathrm{rel}}\) denotes the relative velocity of the incoming protons, and \(f_{pp}(E)\) represents the integrated leptonic phase-space factor including Coulomb distortion effects associated with the emitted positron and is parametrized as $f_{pp}=0.142[1+9.04E] (MeV)$, where E denotes the pp relative energy \cite{Bahcall1969, Schiavilla1998}.


Because the proton--proton fusion process is dominated by the axial-vector component of the weak interaction Hamiltonian, the dominant contribution to the nuclear transition matrix element arises from the one-body axial current operator $A_\mu^{(1)}$ represented as :

\begin{equation}
\langle d,M|A_{\mu}^{(1)}|pp\rangle
=
\delta_{M,\mu}\sqrt{16\pi}\,g_A
\frac{e^{i\delta_0}}{k}
\int_0^{\infty}
dr\,u(r)\chi_0(r;k)
\label{matrix}
\end{equation}

where \(u(r)\) denotes the deuteron radial wavefunction and \(\chi_0(r;k)\) represents the asymptotically normalized proton--proton scattering wavefunction corresponding to the relative momentum \(k\). The quantity \(\delta_0\) is the low-energy \(S\)-wave phase shift, which remains extremely small in the solar-energy region, allowing the approximation

\[
e^{i\delta_0}\approx 1.
\]

and the value of vector-coupling constant, \(g_A\) 

 \[
 g_A = 1.2654 \pm 0.0042
 \]

 is adopted from the weighted average of experimental determinations obtained from neutron beta decay and superallowed Fermi transitions \cite{Adelberger2011}.
Overlap integral $\int_0^{\infty}
dr\,u(r)\chi_0(r;k)$ represents the overlap between pp continuum and bound state deuteron wavefunction.  It is essential to calculate the interaction potential precisely, as it governs the forces between nucleons across a wide range of inter-nucleon distances. Therefore, a realistic interaction potential must simultaneously reproduce the short-range repulsive core, an intermediate attractive nuclear region, and the correct asymptotic behavior in order to describe scattering observables reliably \cite{Machleidt2001,stoks1994construction}.
 

\subsection*{Determination of the Interaction Potential}

In principle, the NN interaction is determined through a direct (forward) approach by solving the Schrödinger equation for a given interaction potential. However, such methods depend intrinsically on prior assumptions regarding the analytical structure of the interaction and may therefore introduce model-dependent ambiguities \cite{ Machleidt2001,Wiringa1995}. For this reason, the present work adopts an inverse-scattering framework \cite{Newton1982,Chadan1989} in which the parameters of the reference potential are determined from scattering observables rather than derived directly from first principles \cite{Epelbaum2009,Carlson2015}. Within this approach, scattering phase shifts play a central role, as they encapsulate the essential information about the interaction over a range of scattering energies and partial waves. 
This procedure resembles a data-driven optimization framework closely aligned with modern machine learning methodologies, emphasizing optimization, uncertainty quantification, and physics-informed learning. \cite{Karniadakis2021,Baker2024}.



The present work employs three smoothly joined Morse functions within the reference potential approach framework \cite{selg2006reference}. The inner component describes the short-range repulsive behavior, the intermediate component reproduces the attractive nuclear interaction, and the outer component governs both nuclear and asymptotic region. This decomposition allows the phenomenological potential to incorporate the distinct physical characteristics of each interaction regime while maintaining sufficient flexibility for inverse scattering analysis.


The parameters of the phenomenological interaction potential are not interpreted independently as direct physical observables. Instead, the physical characteristics of the nucleon–nucleon interaction are determined by the total interaction potential constructed from the smoothly joined functions. Thus, quantities associated with the interaction are inferred from the behavior of the complete potential rather than from any single parameter independently. The parameter values are therefore determined phenomenologically from experimental scattering observables, particularly the phase shifts corresponding to different partial waves.




Following the methodology adopted in recent inverse scattering studies \cite{sastri2024constructing,awasthi2024high,sharma2025novel,sharma2026genetic}, the interaction potential using Morse functions is represented as
\[
V_i(r)=V_i \pm D_i \left( e^{-2\alpha_i (r-r_i)} - 2e^{-\alpha_i (r-r_i)} \right)
\]

where $D_i$ represents the depth of the potential, $\alpha_i$ controls its range and shape, and $r_i$ denotes the equilibrium position.
Total interaction potential, constructed from smoothly joined Morse functions, is written as:
\[
V(r)=
\begin{cases}
V_0(r), & r \leq x_1 \\
V_1(r), & x_1 < r < x_2 \\
V_2(r), & r \geq x_2
\end{cases}
\]

To ensure physical consistency and numerical stability, continuity and differentiability conditions are imposed at the matching points $x_1$ and $x_2$. These conditions require both the potential and its derivative to remain continuous across adjacent regions,
\[
V_i(x_i)=V_{i+1}(x_i),
\]

\[
\left.\frac{dV_i}{dr}\right|_{x_i}
=
\left.\frac{dV_{i+1}}{dr}\right|_{x_i}.
\]
The imposition of these boundary conditions reduces the number of independent parameters from 14 to 10, while simultaneously preventing unphysical discontinuities in the interaction potential.

For a charged system, as in proton–proton scattering, the Coulomb interaction must also be incorporated consistently. Since the Coulomb force is inherently long-range and an accurate determination of the scattering phase shift requires the potential to approach zero as r $\rightarrow \infty$, ensuring the correct asymptotic behavior of the wavefunction.
Unlike conventional approaches that require the separate introduction of additional screened Coulomb ansätze, the present formulation possesses the important advantage that the electromagnetic interaction emerges intrinsically within the Reference Potential Approach itself through three smoothly connected Morse functions. 
In particular, the third Morse component is taken with an inverted sign, thereby accounting for the asymptotic repulsive behavior while maintaining a smooth transition between the nuclear and electromagnetic regions. This treatment avoids the inclusion of artificial screening functions and preserves the continuity of the interaction over the full radial domain. The interaction parameters are determined through the Phase Function Method (PFM), originally introduced by \cite{Calogero1958}, which provides a direct relation between the interaction potential and the scattering phase shifts. In this formalism, the second-order radial Schrödinger equation is transformed into a first-order nonlinear differential equation governing the phase function, thereby circumventing the numerical instabilities associated with direct integration of oscillatory scattering wavefunctions.

For single-channel elastic scattering, the phase function $\delta_l(k,r)$ corresponding to a partial wave $\ell$ satisfies the nonlinear Riccati-type differential equation

\begin{equation}
\frac{d\delta_l(k,r)}{dr}
=
-\frac{V(r)}{k}
\left[
\cos\delta_l(k,r)\,\hat{j}_l(kr)
-
\sin\delta_l(k,r)\,\hat{\eta}_l(kr)
\right]^2 
\label{phaseeq}
\end{equation}

where $\hat{j}_l$ and $\hat{\eta}_l$ are the Riccati--Bessel and
Riccati--Neumann functions, respectively, and k represents the relative momentum of the system.

 At the low energies relevant to proton–proton fusion, the dominant contribution arises from the $\ell=0$ partial wave. In this case, the phase equation simplifies to

\[
\frac{d\delta_0(k,r)}{dr}
=
-\frac{V(r)}{k}
\sin^2 \left( kr + \delta_0(k,r) \right)
\]
The phase function is initialized through the condition $\delta_l(k,0)=0$, and its asymptotic value at large radial distance yields the physical scattering phase shift. The PFM therefore provides a numerically stable and computationally efficient method for extracting scattering observables directly from the interaction potential.

For systems involving tensor interactions or non-central forces, different partial waves become coupled, leading to a multichannel scattering problem. In such cases, the PFM is generalized to a coupled nonlinear first-order differential equation involving two phase shifts corresponding to $L=J+1$ and $L=J-1$ and a mixing parameter \cite{babikov1967phase}. For a given total angular momentum J, the coupled channels $L=J\pm1$ are described through the phase functions as given in \cite{babikov1967phase,sharma2025novel}

For the neutron–proton interaction, the coupled phase equations are solved to determine the effective interaction potentials associated with the coupled 
$^3S_1$
 and 
$^3D_1$
 channels of the deuteron system. The formalism incorporates both the diagonal nuclear interaction terms and the off-diagonal tensor coupling contributions responsible for the mixing between the S- and D-state components. In contrast to the proton–proton system, the neutron–proton is neutral particle interaction and therefore does not contain any Coulomb repulsion or tunneling barrier. Consequently, within the Reference Potential Approach, the third Morse component is retained with a positive sign, allowing the interaction potential to remain purely attractive at large radial distances \cite{awasthi2024high}. This provides a smooth and reliable description of the bound deuteron system within the same unified potential framework. 

The determination of the interaction potential from scattering data constitutes an inverse problem in which the underlying model parameters are not known a priori and must instead be inferred from experimentally observed quantities. 
This formulation is conceptually analogous to machine-learning-based optimization problems. The interaction potential, characterized by a finite set of adjustable parameters, plays the role of the model, while the phase function method acts as the forward operator that maps these parameters to physically observable quantities such as scattering phase shifts. The phase-shift data therefore act as the reference dataset against which the theoretical predictions are evaluated. The optimization procedure can then be interpreted as a training-like process in which the potential parameters are iteratively updated to minimize the Mean Squared Error (MSE) between theoretical calculations and experimental observations. However, unlike purely data-driven machine learning approaches, the present framework remains explicitly constrained by the underlying physical equations governing the scattering process. In this sense, the present approach is more appropriately described as a physics-informed inverse optimization framework employing evolutionary optimization techniques for parameter determination \cite{Karniadakis2021, Goldberg1989, Deb2001}.

This optimization procedure involves the iterative adjustment of model parameters through the minimization of a well-defined cost function. The use of a genetic algorithm further strengthens this analogy, especially for high-dimensional and strongly non-linear optimization landscapes in which conventional gradient-based local optimization methods often become trapped in local minima. In the present framework, the genetic algorithm performs a global exploration of the parameter space and naturally generates multiple near-optimal solutions capable of reproducing the experimental phase-shift data with comparable accuracy.
This ensemble of solutions is not merely a limitation but an advantage, as it enables a systematic analysis of parameter correlations and the propagation of uncertainties to derive observables such as the astrophysical S-factor. Consequently, the present methodology extends beyond conventional parameter fitting and constitutes a physics-informed inverse optimization framework constrained explicitly by the underlying scattering equations and experimental observables.

Leave-one-out cross-validation (LOOCV) \cite{berrar2019cross} is implemented here as a systematic procedure to quantify the sensitivity of the extracted interaction potential to the input data. In this approach, one data point (corresponding to a specific energy) is removed at a time from the full dataset, and the optimization is repeated using the remaining 10 data points. Since the number of data points equals the number of unknown parameters (10 each), the system of equations is exactly determined, yielding a unique and exact solution. This exact determination of the parameters ensures that the solution is mathematically precise, with no ambiguity remaining in the parameter space. Consequently, the uncertainties associated with the extracted parameters are both rigorous and physically consistent. The dataset consists of 11 energy points, covering the S-wave in the pp system and the S-, D-, and tensor wave components in the np system. This procedure is repeated 11 times, each time omitting one energy point from the dataset, thereby yielding 11 distinct sets of optimized potential parameters, each corresponding to a slightly different data configuration.
Rather than using LOOCV purely as a predictive validation tool, it is employed here to analyze the stability of the extracted parameters. The ensemble of parameter sets obtained from all iterations is then statistically analyzed by computing their average values and corresponding deviations. This allows one to quantify how variations in the input data propagate into the potential parameters, thereby providing a direct measure of the sensitivity and uncertainty associated with the fitting procedure.
Such an analysis is particularly important in the present inverse problem, where parameter correlations and non-uniqueness can arise. 

\section*{Determination of Wavefunction}

Once the interaction potential is optimized, it is subsequently employed in the construction of both the proton–proton scattering wavefunction and the deuteron bound-state wavefunction required for the calculation of the astrophysical S-factor. The proton–proton scattering state is obtained within the framework of the Phase Function Method, ensuring consistency between the extracted phase shifts and the corresponding scattering solution. The deuteron bound-state wavefunction is determined through the coupled-channel treatment of the $^3S_1 - ^3D_1$ system, which incorporates the tensor-induced S–D mixing of the deuteron. The detailed formalism and numerical procedure employed for the construction of both the scattering and bound-state wavefunctions are presented in the following sections.

\textbf{ pp-wavefunction}

The proton--proton scattering wavefunction is constructed using the Phase Function Method (PFM) developed by Vladimir Babikov \cite{babikov1967phase}, in which the radial Schr\"odinger equation is transformed into a first-order nonlinear differential equation for the phase function. In this method, the scattering information is obtained directly from the radial evolution of the phase generated by the interaction potential, avoiding the explicit construction of linearly independent oscillatory solutions of the Schr\"odinger equation.

For a central interaction, the radial Schr\"odinger equation is written as

\begin{equation}
\chi_l''(r)+\left[k^2-\frac{l(l+1)}{r^2}-V(r)\right]\chi_l(r)=0,
\end{equation}

where \(k\) is the relative momentum and \(V(r)\) denotes the interaction potential. In the phase-function method, the solution of the radial Schrödinger equation is expressed in terms of the Riccati--Bessel and Riccati--Neumann functions together with a phase function as

\begin{equation}
\chi_l(r)=A_l(r)\left[\cos\delta_l(r)\,\hat{j}_l(kr)-\sin\delta_l(r)\,\hat{\eta}_l(kr)\right],
\end{equation}

where \(\hat{j}_l(kr)\) and \(\hat{\eta}_l(kr)\) are the Riccati--Bessel and Riccati--Neumann functions, respectively. The quantity \(\delta_l(r)\) is the phase function and \(A_l(r)\) is the amplitude function. The phase function describes the phase shift generated by the interaction potential up to the radial point \(r\), while the amplitude function describes the radial evolution of the wavefunction amplitude.

Using the representation of the radial wavefunction in terms of the Riccati functions, the second-order radial Schrödinger equation reduces to a first-order nonlinear Riccati-type differential equation for the phase function, as given in Eq.~(\ref{phaseeq}).

The corresponding amplitude function satisfies the equation

\begin{equation}
\frac{dA_l(r)}{dr}
=
-\frac{1}{k}A_l(r)V(r)
\left[
\cos\delta_l(r)\,j_l(kr)
-
\sin\delta_l(r)\,n_l(kr)
\right]
\left[
\sin\delta_l(r)\,j_l(kr)
+
\cos\delta_l(r)\,n_l(kr)
\right].
\end{equation}

The simultaneous integration of these equations completely determines the scattering wavefunction. Since the phase equation is first order and the phase function varies smoothly with radial distance, the method is numerically stable and particularly suitable for low-energy scattering calculations.

In the present work, the proton-proton interaction potential is obtained through an inverse procedure. The Coulomb contribution is not introduced separately through an explicit external Coulomb term. Instead, the optimized interaction potential effectively incorporates both nuclear and electromagnetic contributions within a unified interaction model. Consequently, the scattering wavefunction is determined entirely from the optimized effective interaction potential through the phase-function equations, removing the requirement of an independently imposed pure Coulomb wavefunction and thereby providing a unified and self-consistent description of the proton--proton system.

At sufficiently large radial distances, where the interaction potential becomes negligible,

\begin{equation}
V(r)\rightarrow 0,
\end{equation}





Since low-energy proton--proton fusion is highly sensitive to the asymptotic behavior of the scattering state, the use of the optimized interaction potential within the PFM framework provides a consistent determination of the scattering observables and overlap integrals without introducing additional external Coulomb corrections.

In the present formulation, the Coulomb interaction is not introduced explicitly as an independent term in the interaction potential, nor are standard Coulomb wavefunctions employed in the construction of the scattering state. Instead, the inverse interaction potential is constructed such that the nuclear and electromagnetic contributions are incorporated effectively within a unified framework. Consequently, the scattering wavefunction is generated entirely from the optimized interaction potential within the phase function formalism, thereby preserving the internal consistency of the approach.

\textbf{ Deuteron Wavefunction}

 The deuteron bound-state wavefunction is obtained through the direct solution of the coupled radial Schrödinger equations \cite{Wiringa1995,Machleidt2001} corresponding to the $^3S_1-^3D_1$
 channels. Owing to the non-central tensor component of the nucleon–nucleon interaction, the deuteron ground state cannot be described as a purely central S-wave configuration. Instead, the tensor force couples the 
$^3S_1$ and $^3D_1$ partial waves, producing an admixture of the D-state component within the deuteron bound state. Consequently, the deuteron must be treated within a coupled-channel framework in which both the diagonal interactions and the off-diagonal tensor interaction are incorporated simultaneously. This coupled-channel structure is essential for reproducing the physical properties of the deuteron, including its binding energy, and asymptotic behavior.

Within this formalism, the coupled radial Schrödinger equations \cite{Machleidt2001} for the neutron–proton system are written as

\begin{equation}
\left[
-\frac{2\mu}{\hbar^2}\frac{d^2}{dr^2}
+ V_S(r)
\right] u(r)
+ V_T(r) w(r)
= E u(r),
\label{eq:cou1}
\end{equation}

\begin{equation}
\left[
-\frac{2\mu}{\hbar^2}\frac{d^2}{dr^2}
+ V_D(r)
+ \frac{6}{r^2}\frac{2\mu}{\hbar^2}
\right] w(r)
+ V_T(r) u(r)
= E w(r).
\label{eq:cou2}
\end{equation}

where u(r) and w(r) denote the radial S- and D-state wavefunctions, respectively, $V_S(r)$ and $V_D(r)$ represent the diagonal interaction potentials for the S- and D-channels, $V_T(r)$ is the tensor coupling interaction, $\mu$ is the reduced mass of the neutron–proton system, and E corresponds to the deuteron binding energy. The centrifugal contribution associated with the D-state appears explicitly through the $\frac{l(l+1)}{r^2}$
 term with $\ell=2$.

In the present work, the diagonal S-state interaction, the D-state interaction, and the tensor coupling interaction are represented independently through piecewise Morse-type functions within the Reference Potential framework, where the corresponding interaction potentials are obtained by solving the coupled nonlinear phase equations.

The deuteron bound-state wavefunction is obtained by solving the coupled-channel Schrödinger equations for the $^3S_1$
 and 
$^3D_1$
 partial waves using the optimized interaction potential. The resulting bound-state solution consists of the radial wavefunctions u(r) and w(r) corresponding to the S- and D-state components of the deuteron, respectively. The coupled-channel wavefunctions are subsequently normalized according to

\[
\int_{0}^{\infty} \left[u^{2}(r)+w^{2}(r)\right] dr = 1
\]

thereby ensuring the proper normalization of the deuteron bound state.

The normalized bound-state wavefunctions obtained from this coupled-channel procedure are subsequently employed in the evaluation of the overlap integral governing the proton–proton fusion transition matrix element and the corresponding astrophysical S-factor.


The cross-section calculated from Eqn. \ref{pcs} exhibits strong energy dependence arising from the Coulomb interaction.
Therefore, direct determination of cross-section at astrophysical energy becomes difficult due to the amplification of experimental uncertainties. By expressing the reaction cross section in terms of the astrophysical S-factor, the dominant energy dependence associated with the Coulomb barrier penetration and the (1/E) behavior is removed, yielding a slowly varying function of energy \cite{Iliadis2007}. Accordingly, the astrophysical \(S\)-factor is calculated using

\begin{equation}
S(E)
=
\sigma(E)\,
E\,
\exp
\left[
2
\int_{r_1}^{r_2}
\sqrt{
\frac{2\mu}{\hbar^2}
\left(
V_{\mathrm{eff}}(r)-E
\right)
}
\,dr
\right].
\label{sf}
\end{equation}

Here, \(r_1\) and \(r_2\) denote the classical turning points corresponding to the effective interaction potential \(V_{\mathrm{eff}}(r)\), while \(\mu\) represents the reduced mass of the proton--proton system. In the present work, instead of using the conventional Coulomb approximation, we have utilized the fundamental WKB action integral evaluated using the effective interaction potential derived through the Reference Potential Approach. This formulation preserves the low-energy tunneling dynamics encoded in the effective interaction potential and therefore provides a more physically consistent description of the proton-proton fusion process in comparison with simplified asymptotic Coulomb approximations.
S-factor further facilitates a reliable extrapolation to the astrophysically relevant low-energy region, from which the corresponding reaction cross section can be reconstructed. Cross-section at low energies are determined by extrapolating the values of the S-factor using a Taylor series expansion. Therefore, the corresponding values of taylor series are determined using quadratic extrapolation as
\begin{equation}
 f(E)=a+bE+cE^2   
 \label{ts}
\end{equation}

where (a), (b), and (c) are parameters obtained from the fit. In this (a), (b), and (c) represent S(0), $S'(0)$, and $S''(0)$, respectively.
One fundamental question that now arises is how far the extrapolation must be extended toward lower energies to reach the astrophysically relevant region. The extrapolation is carried out up to the Gamow window, which corresponds to the energy at which the reaction probability attains its maximum value. The Gamow energy window defines the astrophysically relevant energy range in which the product of the reaction cross section and the Maxwell–Boltzmann distribution contributes most significantly to the thermonuclear reaction rate and stellar energy generation. The thermonuclear reaction rate is given by
\begin{equation}
\langle \sigma v \rangle =
\left( \frac{8}{\pi \mu} \right)^{1/2}
\frac{1}{(k_B T)^{3/2}}
\int_{0}^{\infty}
\sigma(E)\,E\,e^{-E/(k_B T)}\,dE.
\end{equation}

using Eqn\ref{sf} it reduces to
\begin{equation}
\propto
\int_{0}^{\infty}
S(E)\exp\left(-\frac{E}{kT}\right)
\exp\left[
-\frac{2}{\hbar}
\int_{r_1}^{r_2}
\sqrt{2\mu\left(V(r)-E\right)}\,dr
\right]
\, dE 
\label{ge}
\end{equation}

here, integrand is the product of Maxwell–Boltzmann distribution, S(E) and Gamow factor, whose maximum defines the Gamow energy peak \cite{Iliadis2007}.



\section{Results and Discussion}
In this section, we construct the effective nucleon-nucleon interaction potentials for both the np and pp system using the Reference Potential Approach (RPA). A Leave-One-Out Cross-Validation (LOOCV) analysis is performed to assess the energy dependence and robustness of the extracted potentials. The sensitivity of the astrophysical S-factor to the extracted interaction potentials is assessed through calculations of the overlap integral and the radiative capture cross section. In the present work, the pp scattering wavefunction is obtained directly using the Phase Function Method (PFM), rather than by solving the Schrödinger equation and matching the solution to the asymptotic Coulomb wavefunction (valid only for the bare Coulomb interaction). This ensures consistency between the interaction potential and the corresponding scattering wavefunction.
As an initial validation, radiative capture cross sections are calculated at the available laboratory energies using both the AV18 interaction and the one obtained using the Reference Potential Approach to assess the accuracy of the corresponding scattering wavefunctions. The close agreement between the overlap integral and radiative capture cross section indicates that the latter provides an accurate description of the scattering wavefunctions for the subsequent S-factor calculations. Next, the astrophysical S-factor is calculated over the energy range of 0.2 - 0.1 MeV, corresponding to the onset of the Coulomb barrier. The resulting S-factor is extrapolated to 0.01 MeV, from which the radiative capture cross section is determined and compared with that calculated directly using the theoretical capture cross-section expression. 
Following the procedure adopted in Ref. \cite{Schiavilla1998}, where the zero-energy astrophysical S-factor is obtained by extrapolating the values calculated in the 1 - 5 keV energy range, we evaluate the radiative capture cross section and the corresponding S-factor over the same energy interval using the present formalism. The results show that employing the Gamow factor based on the bare Coulomb approximation leads to an overestimation of the astrophysical S-factor. 
Furthermore, the interaction potential obtained through the Reference Potential Approach (RPA) provides the advantage of enabling the calculation of the radiative capture cross section and the corresponding astrophysical S-factor at energies as low as $10^{-4}$ MeV. The resulting S-factor is then extrapolated to zero energy using a neural network, rather than the conventional polynomial fitting based on a Taylor-series expansion.
The significance of performing these calculations and the insights they provide into the determination of the zero-energy astrophysical S-factor are discussed in detail below.
\\

The interacting protons in the proton-proton weak capture reaction possess extremely small relative energies inside stellar interiors. Therefore, the probability of tunneling through the Coulomb barrier is strongly suppressed, resulting in a small fusion cross-section. At laboratory energies, the cross section is calculated using Eqn.~\ref{pcs}. The corresponding astrophysical S-factor is then determined using actual Gamow penetration factor that involves the calculation of an integral 
\begin{equation}
 I=   \frac{2}{\hbar}\int_{r_1}^{r_2}\sqrt{2\mu(V(r)-E)}\,dr
 \label{wkbi}
\end{equation}
where V(r) is the effective interaction potential obtained from reference potential approach.
It is evident from the results presented in Table \ref{comp} that reducing the WKB tunnelling integral to the ($2\pi\eta$) factor by considering only the bare Coulomb potential leads to an overestimation of the calculated values.

\begin{table}[h!]
\centering
\caption{Comparison of the Sommerfeld Parameter and the WKB Integral $I$ for different energies}
\begin{tabular}{ccccc}
\hline
$E_{lab}(MeV)$     & $2\pi\eta$  & $I$     & $e^{(2\pi\eta)}$ & $e^{(2I)}$ \\ \hline
0.1   & 3.1405  & 0.3448  &23.1147      & 1.9928  \\ 
0.01  & 9.9310  & 0.5402 & 20558.6482   & 2.9458  \\ 
0.001 & 31.4047 & 0.6022 & $4.3540\times 10^{13}$  & 3.3346  \\ \hline
\end{tabular}

\label{comp}
\end{table}

A significant discrepancy is observed between the values obtained using the Sommerfeld factor and those derived from the fundamental WKB integral. The effect of this difference will be reflected in the resulting Gamow peak energy and S-factor at low energies.


\textbf{Determination of Interaction Potential and Wavefunction:}\\
To overcome the limitations arising due to the inclusion of the bare Coulomb potential, we have obtained the interaction potential for both the np and pp systems using reference potential approach. The inverse potential was constructed using a genetic algorithm-based optimization technique in conjunction with the PFM.
A key advantage of this methodology is its ability to naturally incorporate both the nuclear and Coulomb interactions without requiring an explicit form for each contribution. Leave-one-out cross-validation was performed to assess the sensitivity of the optimized parameters to individual data points. In the present analysis, eleven different sets of proton-proton and neutron-proton interaction potentials together with the corresponding continuum and bound-state wavefunctions are generated using the Leave-One-Out Cross-Validation (LOOCV) procedure within the Reference Potential Approach framework.
The resulting continuum and bound-state wavefunctions are then used consistently in the evaluation of the overlap integral governing the proton-proton fusion transition amplitude. 
A small spread in the obtained parameter sets indicates that the potential is robust and weakly sensitive to
individual data points, whereas larger deviations reflect stronger dependence on the input dataset. This ensemble-based approach therefore provides a systematic way to assess the reliability of the constructed
interaction potential and its impact on subsequent observables. \\

\begin{figure}
    \centering
    \includegraphics[width=0.5\linewidth]{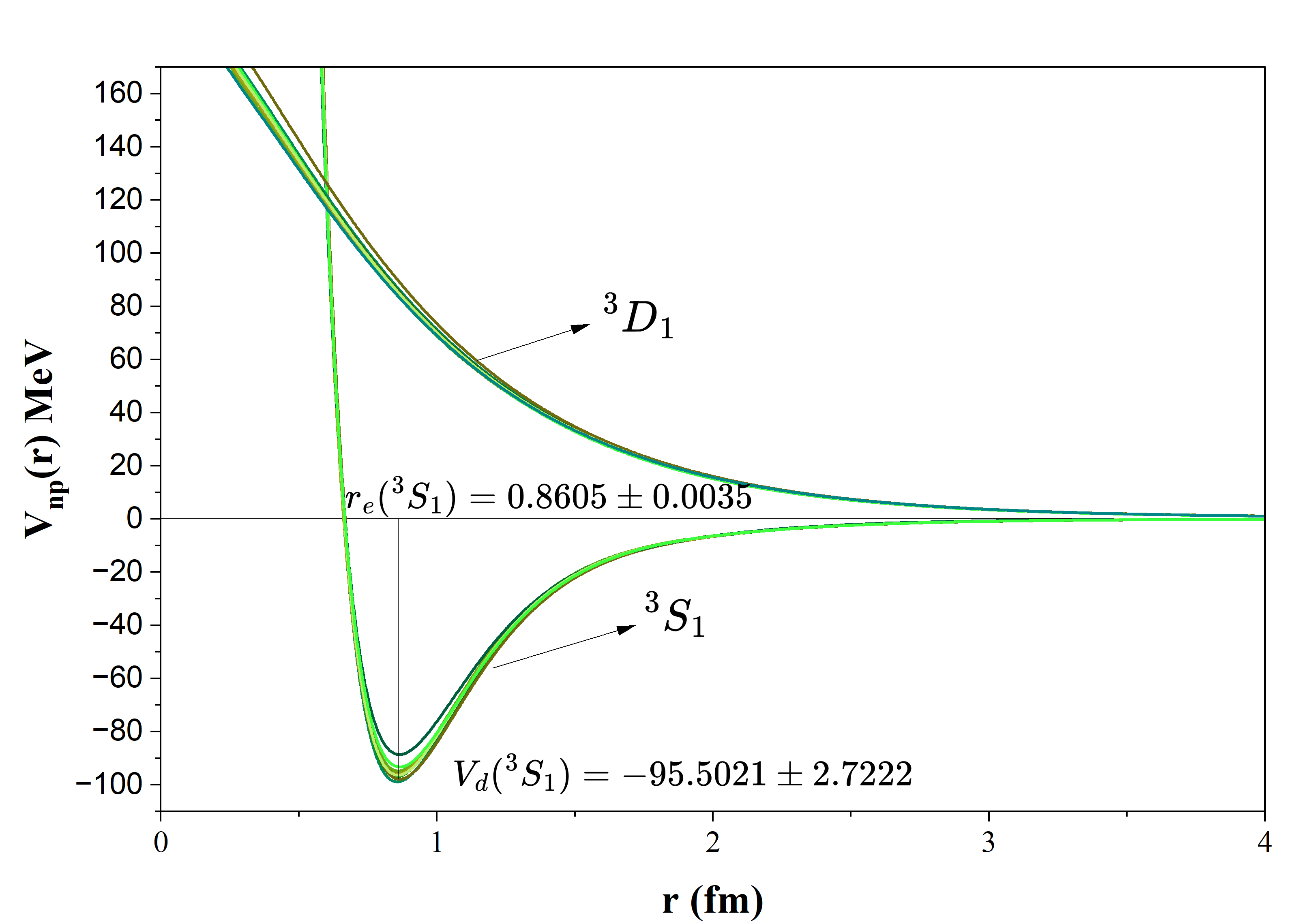}
    \caption{Optimized neutron--proton interaction potential obtained in the present work, along with the corresponding average values and uncertainties of the physical parameters are presented.}
    \label{nppot}
\end{figure}

\textbf{n-p interaction}
\\
The obtained eleven sets of neutron–proton interaction potentials are shown in Fig.~\ref{nppot}. The physical parameters presented in the plots are obtained from the total interaction potential, which is constructed using the complete set of model parameters. Therefore, these physical parameters are determined by the combined effect of all the model parameters rather than by any single model parameter. For the neutron–proton interaction, the average depth of the $^3S_1$ potential is obtained as $V_d(^3S_1)= -96.5858\, \pm \,2.7220~ MeV$. The associated uncertainty indicates a noticeable spread in the calculated values of the $^3S_1$ potential depth. Since these optimized potentials are obtained using different energy datasets, the observed variation in the potential depth reflects the sensitivity of the interaction potential to the input scattering data. This behavior demonstrates that even small changes in the experimental energy points can produce measurable variations in the extracted potential parameters, indicating the underlying energy dependence and non-uniqueness of the inverse scattering problem. In contrast, the relatively small uncertainty in the equilibrium distance $r_e(^3S_1)=0.8605\, \pm \,0.0035~ fm$ demonstrates the stability and consistency of the obtained interaction potential within the Reference Potential Approach. The obtained $^3S_1$ channel exhibits the attractive behavior necessary for the deuteron bound-state formation, while the $^3D_1$ channel shows a comparatively repulsive character consistent with the negative trend of the corresponding phase shifts.

\begin{figure}
    \centering
    \includegraphics[width=0.5\linewidth]{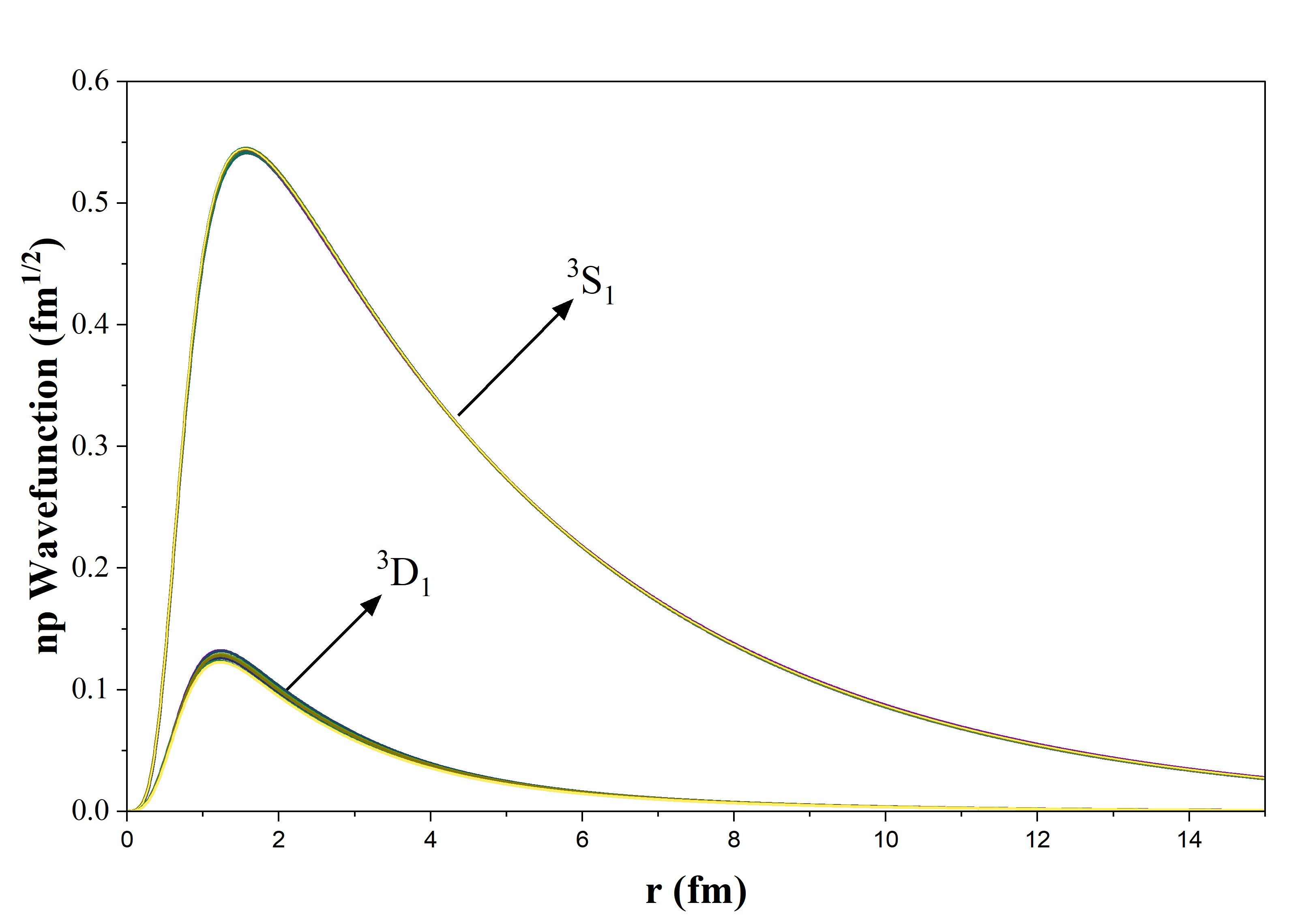}
    \caption{Normalized deuteron bound-state radial wavefunctions obtained from the coupled-channel solution of the neutron–proton Schrödinger equations using the optimized interaction potential. The upper curve corresponds to the dominant $^3S_1$ component, represented by the radial wavefunction $u(r)$, while the lower curve corresponds to the tensor-coupled $^3D_1$ component, represented by $w(r)$.}
    \label{bsw}
\end{figure}

The corresponding bound-state wavefunctions for the $^3S_1$ and $^3D_1$ states obtained from each optimized interaction potential are presented in Fig.~\ref{bsw}. This procedure ensures that the uncertainties originating from the scattering data are consistently propagated into the deuteron bound-state solution. 
The radial coordinate is discretized on a uniform grid extending to sufficiently large distances to ensure convergence of the bound-state solutions. The coupled Schrödinger equations, Eqs.~(\ref{eq:cou1}) and (\ref{eq:cou2}), are then discretized using the Numerov method \cite{Johnson1977}, resulting in a generalized matrix eigenvalue problem for the bound-state energies and wave functions.
The Hamiltonian matrix contains the kinetic-energy contribution together with the diagonal interaction potentials and the tensor coupling terms mixing the S- and D-state channels.
The resulting generalized eigenvalue problem is solved numerically to obtain the eigenvalues and eigenvectors of the coupled neutron–proton system. 
The deuteron binding energy is therefore determined directly from the eigenvalue spectrum of the coupled-channel system.
The calculations performed using the ensemble of optimized interaction potentials yield a deuteron binding energy in excellent agreement with the experimental value. The statistical analysis of the obtained binding energies gives an average value of $E_B=-2.22457 \,\pm \,0.00843~ MeV $,
where the quoted uncertainty corresponds to the standard deviation obtained from all optimized potential configurations. The close agreement between the calculated average binding energy and the experimentally observed deuteron binding energy $B_d(exp)=-2.224575~ MeV$ \cite{Tilley2002} demonstrates the reliability and consistency of the coupled-channel numerical framework, as well as the accuracy of the optimized neutron–proton interaction potentials employed in the present work.

\begin{figure}
    \centering
    \includegraphics[width=0.5\linewidth]{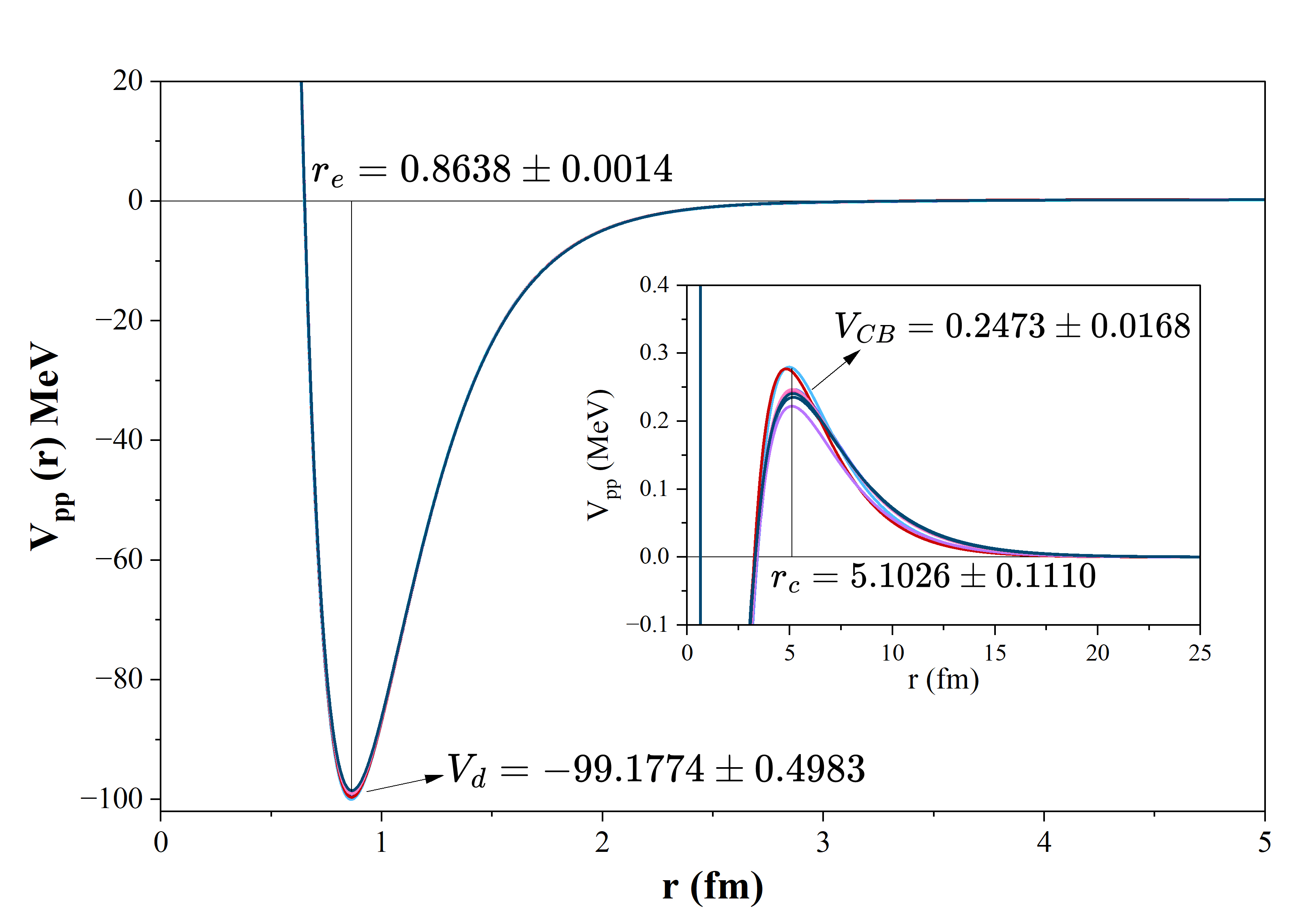}
    \caption{Optimized proton-proton interaction potential obtained in the present work. The figure shows the potential depth $V_d$, Coulomb barrier height $V_{CB}$, and the corresponding radial positions $r_e$, together with their associated uncertainties.}
    \label{pppot}
\end{figure}

\textbf{p-p interaction}\\
The obtained eleven sets of proton-proton interaction potentials are shown in Fig.~\ref{pppot}. For proton–proton interaction, uncertainties associated with the extracted physical parameters are smaller, indicating greater parameter stability and a weaker sensitivity of the interaction potential to the selected energy points used in the analysis. The depth of the attractive potential is $V_d = -99.1774 \,\pm \,0.4983~ MeV$ at the equilibrium distance of $r_e = 0.8638\, \pm \,0.0014 fm$. At larger radial distances, the interaction develops a repulsive Coulomb barrier whose maximum height is obtained as $V_{CB} = 0.2473 \,\pm \, 0.0168 ~MeV$ at $r_c = 5.1026  \,\pm \, 0.1110~ fm$. The smooth transition from the short-range attractive nuclear region to the long-range repulsive electromagnetic region demonstrates the effectiveness of the reference potential approach in providing a unified description of the proton–proton interaction.

LOOCV analysis generates 11 independent sets of optimized interaction parameters corresponding to different subsets of the scattering phase-shift data. For each of these optimized interactions, the corresponding proton–proton scattering wavefunction is reconstructed within the phase function formalism by solving the associated phase and amplitude equations. An important feature of the phase-function method is that it eliminates the need to match the numerical wavefunction to the conventional Coulomb wavefunctions, which are derived for the bare Coulomb potential. This follows from the fact that all relevant interactions are incorporated within a unified potential, without explicitly treating the Coulomb interaction as a separate contribution.

\begin{figure}[h]
    \centering
    \includegraphics[width=0.5\linewidth]{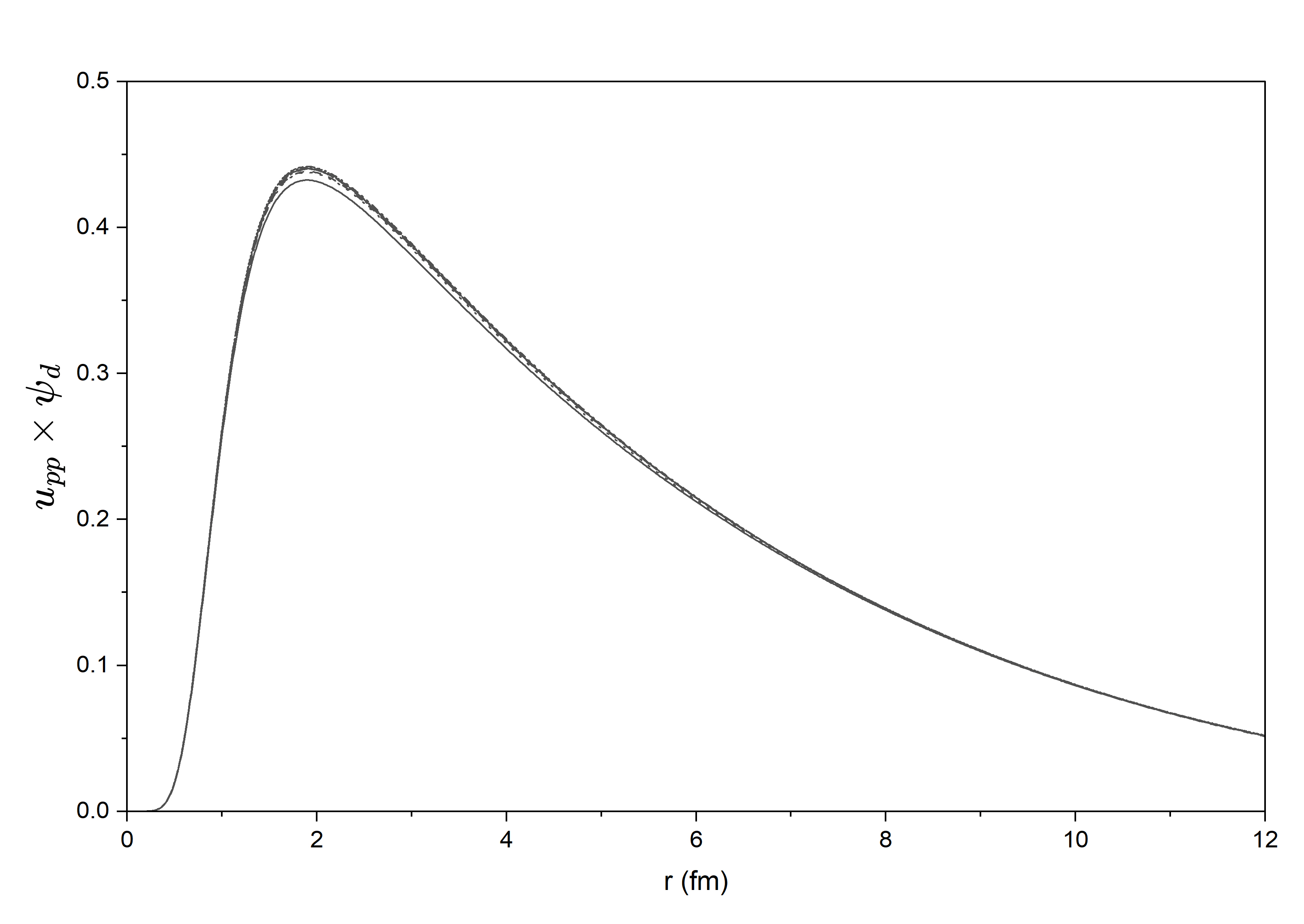}
    \caption{Overlap of the proton–proton and deuteron wave functions at E=1 MeV for 11 potentials obtained using LOOCV.}
    \label{op}
\end{figure}

The overlap between two wave functions for 11 np and pp potentials as a function of radial distance is plotted in Fig.~\ref{op} for $E = 1 ~MeV$. Figure shows that, although potential parameters exhibit significant variations, the resulting variation in the value of overlap is relatively small.
Although computationally intensive, such a treatment is essential for obtaining a systematic and self-consistent estimation of uncertainties in the final observables. In the present work, each calculated result is presented in form $x \,\pm \,\delta x$, where $ x$ is the mean value obtained from eleven interaction potentials, and $\delta x$ is the corresponding standard deviation, representing the uncertainty in calculated quantity. \\




The traditional calculation based on the bare Coulomb approximation indicates that the evaluation of the astrophysical S-factor is not fully self-consistent. While the overlap integral is calculated using wave functions generated from the chosen interaction potential, the tunnelling probability is represented by the Sommerfeld (Gamow) factor derived by considering only the bare Coulomb potential. A fully self-consistent treatment requires the tunnelling probability to be evaluated using the same interaction potential employed to generate the wave functions.
This approximation is adequate when the astrophysical S-factor is determined from cross sections measured at laboratory energies well above the Coulomb barrier, where tunneling effects are almost negligible. For example, in the pp system, extrapolating data from 1–5 MeV, with the Coulomb barrier being only about 0.3 MeV, would not necessitate an explicit WKB evaluation. However, such an extrapolation would lead to an S(0) value significantly larger than the currently accepted one.
Within the WKB framework, the tunneling probability depends explicitly on the interaction potential and the associated classical turning points. The total interaction potential, comprising both the nuclear and Coulomb components, must be taken into account. For a pure Coulomb potential, the WKB barrier penetrability reduces to $\exp(2\pi\eta)$, involving sommerfeld parameter. Since no finite-range nuclear interaction is included, the inner turning point is effectively taken near the origin, and the tunneling probability is governed predominantly by the outer Coulomb turning point. This treatment completely neglects the nuclear potential. However, when the total potential is evaluated over the relevant range of inter-particle distances by considering its value at each radial point, both classical turning points are correctly obtained. 
Therefore, an accurate determination of the tunneling probability requires numerical evaluation of the total potential to correctly define both turning points of the classically forbidden region. The present analysis ensures that both the pp continuum wavefunction and associated tunneling probability determined consistently from same effective interaction potential. Tunneling probability calculated using the obtained effective potential in the fundamental WKB integral retains the detailed low-energy dynamical information.
Consequently, the calculated overlap integral and astrophysical \(S\)-factor remain fully consistent with the underlying low-energy proton-proton interaction employed in the present analysis.
The overlap integral $\int_{0}^{\infty} dr \, u(r)\chi_0(r;k)$ determines the transition matrix element between the initial proton–proton continuum state and the final deuteron bound state and therefore plays a central role in determining the proton–proton fusion cross section and the corresponding astrophysical S-factor.\\

\textbf{Radiative Capture Calculations in the 1 - 5 MeV Energy Range:}\\
Using the nuclear matrix element contribution, Eqn.~\ref{matrix}, obtained from the axial one-body current together with the corresponding proton–proton fusion cross section, Eqn.~\ref{pcs}, we first calculate the astrophysical S-factor at the lab energies from 1 MeV to 5 MeV in steps of 1 MeV for both RPA and AV18.

\begin{table}[h]
\centering
\caption{Radiative capture cross sections calculated at laboratory energies using the AV18 potential and the potential obtained through the reference potential approach(RPA).}
\label{icomp}
\begin{tabular}{|l|lll|lll|}
\hline
 Potentials$\to$ & \multicolumn{3}{c|}{RPA}      & \multicolumn{3}{c|}{AV18}       \\ \hline
$E_{lab}(MeV)$ &
  \multicolumn{1}{l}{$\delta(degrees)$} &
  \multicolumn{1}{l}{overlap} &
  \multicolumn{1}{l|}{sigma(barn)} &
  \multicolumn{1}{l}{$\delta(degrees)$} &
  \multicolumn{1}{l}{overlap} &
  \multicolumn{1}{l|}{sigma(barn)} \\ \hline
1 & $32.7558\pm 0.4692$ & $2.6245\pm0.0122$ & $4.0697\pm0.0377\times 10^{-25}$ 
& 32.6274 & 2.5049 & $3.7121\times 10^{-25}$  \\ 
2 & $44.7024\pm 0.3348$ &$ 2.5994\pm0.0084$ & $1.4114\pm 0.0090\times 10^{-25}$ 
& 45.5653 & 2.6028 & $1.4171\times 10^{-25}$ \\ 
3 & $50.4117\pm 0.2689$ & $2.3268\pm 0.0067$ & $6.1557\pm 0.0355\times 10^{-26}$ 
&  50.9137 & 2.3876 & $6.4909\times 10^{-26}$  \\ 
4 & $53.2665\pm 0.2011$ & $2.0594\pm0.0072$ & $3.1322\pm 0.0219\times 10^{-26}$
& 53.4541 & 2.1562 & $3.4382\times 10^{-26}$  \\ 
5 & $54.7084\pm0.1454$ & $1.8337\pm 0.0070$ & $1.7769\pm0.0137\times 10^{-26}$
&  54.7165 & 1.9522 & $2.0167\times 10^{-26}$  \\ \hline
\end{tabular}
\end{table}

As shown in Table~\ref{icomp}, the overlap integrals and radiative capture cross sections calculated using the two interaction potentials are in close agreement at laboratory energies. The excellent agreement demonstrates the reliability of the wavefunction obtained using the PFM and confirms its consistency with the conventional approach based on solving the Schrödinger equation followed by asymptotic matching to the Coulomb wavefunctions. The overlap integral and cross section corresponding to the AV18 potential were calculated following the same methodology as that described by Schiavilla \cite{Schiavilla1998}. Since direct experimental measurements are generally not feasible in the low-energy region corresponding to the Gamow window, the reaction cross section at stellar energies is determined using the extrapolated astrophysical $S$-factor. In principle, the extrapolation should be performed using the available laboratory-energy data, where the phase shifts are accurately determined. Therefore, one would expect that extrapolating the astrophysical $S$-factor from these lab energies (Table~\ref{icomp}) would reproduce the corresponding low-energy results. However, this is not the case, as the extrapolated values do not agree with those reported in the literature.
Traditionally, instead of directly extrapolating the astrophysical $S$-factor from the laboratory-energy region, the cross section is first calculated down to the low-energy region using Eqn.~\ref{pcs}. These calculated cross sections at low energies are, however, also subject to uncertainties associated with the extrapolation of phase shifts. The corresponding astrophysical $S$-factor is then evaluated in the low-energy region and subsequently extrapolated to zero energy to determine $S(0)$.

Therefore, next we have calculated  astrophysical S-factor using the effective proton-proton interaction potential obtained through the Reference Potential Approach over the energy region extending from 0.2 MeV up to 0.1 MeV, as shown in Table~\ref{tab:sfactor} using Eqn.~\ref{sf}. Here, the uncertainties listed in the table represent the total error, obtained by combining the uncertainty from the LOOCV analysis 

\begin{table}[h]
\centering
\caption{Astrophysical S-factor values and corresponding uncertainties, calculated using Eqn.~\ref{sf}, based on interaction potentials obtained using the Reference Potential Approach.}
\resizebox{1.0\textwidth}{!}{%
\begin{tabular}{cccccc}
\hline
$E_{lab}$(MeV) & $\delta(degrees)$    &    overlap          &  $e^{2I}$            & $\sigma(E)(barn)$ & $S(E)(MeV barn)$ \\
\hline
0.2            & $13.0312\pm0.5831$   & $1.5131\pm0.0138$   & $1.5648 \pm 0.0408$  & $(1.5146\pm 0.0280)\times10^{-24}$ & $(2.3691\pm0.0337)\times10^{-25}$ \\

0.15           & $11.1181\pm0.5389$   & $1.3268\pm0.0138$   & $1.7383 \pm 0.0456$  & $(1.7928\pm 0.0377)\times10^{-24}$  & 
$(2.3363 \pm 0.0260)\times10^{-25}$ \\

0.1            & $8.9308\pm 0.4690$   & $1.0961\pm0.0129$   & $1.9694 \pm 0.0554$ &  $(2.2479\pm 0.0537)\times10^{-24}$ & $(2.2122 \pm 0.0189) \times10^{-25}$  \\

\hline
\end{tabular}}
\label{tab:sfactor}
\end{table}
The WKB action integral $I$ decreases with increasing incident energy because the width of the classically forbidden region becomes smaller at higher energies, thereby reducing the effective tunneling path through the barrier. 
The widening of the classically forbidden region with decreasing energy increases the WKB action integral, leading to stronger exponential suppression in the transmission probability, i.e $e^{-2I}$. Here, from Table~\ref{tab:sfactor} it is visible that as the WKB action integral increases, the exponential factor $e^{2I}$ increases correspondingly, effectively removing the tunneling suppression from the fusion cross-section.
The effect becomes especially significant at very low energies where the tunneling region is substantially wider.

Similarly for the AV18 potential, the astrophysical S-factor was calculated using Eqn.~\ref{sf}. The obtained S-factor values are presented in Table~\ref {AV18s} both using the WKB integral $I$ and the factor consist of sommerfeld parameter $\eta$. The overlap integral and cross-section values corresponding to the AV18 potential were calculated using the same methodology as described by Schiavilla \cite{Schiavilla1998}.

\begin{table}[h!]
\centering
\caption{Astrophysical S-factor for the AV18 potential calculated using the fundamental WKB integral given by Eqn.~\ref{sf}.}
\label{AV18s}
\begin{tabular}{cccccccc}  
\hline
$E_{lab}$(MeV) & $\delta (degrees)$  & overlap & $\sigma(barn)$        &$exp(2I)$ & $S(I) (MeV barn)$     &  $exp(2\pi\eta)$ & $S(\eta) (MeV barn)$    \\  \hline
0.2            & 6.7418 & 0.9409   & $5.8562\times10^{-25}$ & 1.9856  & $1.1553\times10^{-25}$ &  9.2133    &  $5.3955\times10^{-25}$           \\
0.15           &4.5083 & 0.7170   & $5.2359\times10^{-25}$ & 2.7965  & $1.0911\times10^{-25}$ &  12.9900   &   $5.1011\times10^{-25}$              \\
0.1            &2.3663 & 0.4680   & $4.0985\times10^{-25}$ & 4.9684  & $1.0115\times10^{-25}$ &  23.1147   &   $4.7368\times10^{-25}$             \\   \hline
\end{tabular}
\end{table}

As can be clearly observed from Table~\ref{AV18s}, the $2\pi\eta$ factor calculated using the bare Coulomb potential yields considerably higher values of the astrophysical S-factor. The bare Coulomb approximation accounts only for the electrostatic repulsion between the interacting nuclei and neglects the effect of nuclear interactions and screening corrections. Consequently, it overestimates the tunneling probability and increases the S-factor values. This result highlights the necessity of employing the total nucleon-nucleon interaction potential in the Gamow factor to obtain a more accurate description of low-energy nuclear reaction rates and fusion processes.
Tables~\ref{tab:sfactor} and \ref{AV18s} clearly demonstrate the difference between the phase shifts obtained using the AV18 and RPA potentials. Both potentials reproduce the experimental phase shifts accurately over the laboratory energy range of 1 - 350 MeV. However, at lower energies, the predicted phase shifts are based on an extrapolation of potentials constrained by the available laboratory-energy data and therefore carry the associated extrapolation uncertainties. In the case of the AV18 potential, the phase shift varies rapidly at low energies due to the strong dependence of the Coulomb wave function on the Sommerfeld parameter $\eta$. Consequently, this rapid variation leads to an overestimation of the calculated values. The differences in the phase shifts are reflected in the corresponding values of the overlap integral.

The S-factor obtained using the overlap integral and the WKB tunneling probability is subsequently extrapolated from the calculated low-energy region toward zero energy for comparison with astrophysical and solar-model calculations. Such an extrapolation becomes necessary because direct numerical calculations at zero energy are hindered by the small fusion probability and the associated numerical instabilities associated with the tunneling region.
Typically, in all calculations, the s-factor is extrapolated up to zero energy using Taylor series expansion. The coefficients of taylor series are obtained using second-order polynomial fitting procedures.
Conventionally, such extrapolations are performed using polynomial fitting procedures in which the low-energy \(S\)-factor is represented through linear, quadratic, cubic, or higher-order polynomial expansions \cite{acharya2016uncertainty, acharya2023revisiting}. 
Solar Fusion II \cite{Adelberger2011} states that employing higher-order polynomial fits for the $^3He-{^3He}$ reaction introduces additional free parameters, making the extrapolation increasingly unstable because these parameters are poorly constrained by the available data. Consequently, a quadratic polynomial was chosen as the preferred extrapolation function, providing a stable description of the low-energy S-factor with reduced fitting uncertainty. However, polynomial fitting is fundamentally an approximate mathematical representation whose behavior strongly depends on the chosen polynomial degree. Higher-order polynomial functions often attempt to forcefully fit the numerical data and may introduce artificial curvature or oscillatory behavior in the extrapolated region. The astrophysical S-factor values given in Table~\ref{tab:sfactor} were fitted with a quadratic polynomial to extrapolate up to zero energy, resulting in $S(0)= (2.2219\pm 0.024) \times 10^{-25}~ MeV~ barn$. The quoted uncertainty arises from the variation in the potential parameters obtained through the LOOCV analysis of different energy data sets.  

The importance of the S-factor is that it enables a reliable determination of the fusion cross-section at very low energies relevant to stellar interiors. After removing the Coulomb barrier effect, the S-factor becomes a slowly varying function of energy that can be accurately extrapolated to zero energy. The extrapolated value S(0) is then used to evaluate the low-energy fusion cross-section required for calculating thermonuclear reaction rates and modelling stellar evolution. 

Now, if we need to obtain the cross section at $E = 0.01~MeV$, there are two approaches. The first approach is to calculate the cross section, $\sigma(0.01)$, using the extrapolated value of the S-factor, S(0), in Eqn.~\ref{ts}. Substituting S(0) into Eqn.\ref{ts}, the S-factor at $E = 0.01$, is obtained as: $S(0.01)= 1.7510\times 10^{-25}~MeV~ barn$
 and using 
\begin{equation}
    \sigma(E) = \frac{S(E)}{E}e^{-2I}
\label{cse}    
\end{equation}

cross section is calculated as $ \sigma(0.01)= 1.2174 \times 10^{-23}~barn$. 

Second approach is to obtain the cross section directly from the empirical calculation of the overlap integral and cross section using Eqn.~ \ref{pcs} and \ref{matrix}. Cross section calculated at $E = 0.01$ using tunneling through Coulomb barrier is given as $\sigma(0.01)= 0.7333\times 10^{-23}~barn$.

This shows the large difference between the values obtained from two different procedures.  Hence, we cannot simply rely upon the extrapolated values. As the values obtained using the extrapolation depend upon the data used for fitting and also vary with the choice of the initial and final data point used for the extrapolation. In such an extrapolation, the obtained values consist of multiple errors, firstly due to extrapolation in values of phase shift and then in the extrapolation of S-factor values. Since the phase shifts must be extrapolated to the low-energy region to obtain the scattering wave functions in order to evaluate the overlap integral. Therefore, the uncertainty introduced by this extrapolation is unavoidable and is common to all potential models. Hence, it accumulates large extrapolation errors, which result in such a huge difference in the cross-section values.
Here, too, extrapolation of the $S (I)$ values obtained using the AV18 potential (Table \ref{AV18s}) does not yield the $S(0)$ value reported in the literature. Since the calculated $S(E)$ value at $E=0.1$ MeV is already lower than the corresponding literature value of approximately $4.0\times 10^{-25}~MeV~barn$, the extrapolated value of $S(0)$ is also expected to remain lower than the reported value.\\

\textbf{Calculations at Low Energies (Down to 1 keV):}\\
In most reported studies, the calculations have been extended to energies as low as 1~keV. In this energy region, the radiative capture cross section decreases continuously, while the corresponding phase shift decreases to values of the order of $10^{-10}$(Table~\ref{tab1k}). This behavior is primarily attributed to the strong energy dependence of the Sommerfeld parameter, $\eta \propto 1/\sqrt{E}$, appearing in the Coulomb wave functions, whose influence becomes increasingly dominant in the low-energy region. As the energy decreases, Sommerfeld parameter ($\eta$) becomes increasingly dominant, causing the phase shift to become extremely small. Since the radiative capture cross section depends on the phase shift through the overlap of the scattering and bound-state wavefunctions, the strong Coulomb suppression at such low energies, characterized by the Sommerfeld parameter ($\eta$), results in correspondingly small values of the calculated cross section.

\begin{table}[h]
\caption{Comparison of the astrophysical S-factor and radiative capture cross sections calculated up to 1 keV using the AV18 potential and the interaction potentials obtained via the reference potential approach (RPA) based on the LOOCV analysis.}
\label{tab1k}
\resizebox{\textwidth}{!}{%
\begin{tabular}{|c|ccc|ccc|}
\hline
\multicolumn{1}{|c|}{} & \multicolumn{3}{c|}{RPA} & \multicolumn{3}{c|}{AV18}      \\ \hline
\multicolumn{1}{|c|}{Elab MeV} &
  \multicolumn{1}{c|}{0.002} &
  \multicolumn{1}{c|}{0.005} &
  \multicolumn{1}{c|}{0.01} &
  \multicolumn{1}{c|}{0.002} &
  \multicolumn{1}{c|}{0.005} &
  \multicolumn{1}{c|}{0.01} \\ \hline
$\delta(deg)$            & $1.2155  \pm 0.0748 $    & $1.9242     \pm 0.1179 $    & $2.7266     \pm 0.1657$     & $1.0831\times10^{-8}$ & $3.8057\times10^{-5}$ & 0.0023 \\
overlap             & $0.1575\pm 0.0025$     & $0.2493     \pm 0.0038$     & $0.3524     \pm 0.0052  $   & $1.1678\times10^{-5}$ & $8.7086\times10^{-4}$ & 0.0081 \\
$\sigma (barn)$           & $1.6408 \pm 0.0522\times10^{-23}$ & $1.0404 \pm 0.0319\times10^{-23}$ & $7.3494 \pm 0.2208\times10^{-24}$ & $9.0213\times10^{-32}$ & $1.2691\times10^{-28}$ & $3.9058\times10^{-27}$ \\
$S(E) MeV~barn $          & $5.2025 \pm 0.1109\times10^{-26}$ & $7.8788 \pm 0.1496\times10^{-26}$ & $1.0572 \pm 0.0176\times10^{-25}$ & $3.9757\times10^{-25}$ & $3.9896\times10^{-25}$ & $4.0149\times10^{-25}$\\ \hline
\end{tabular}}
\end{table}
Consequently, the astrophysical $S$-factor obtained from these cross sections is strongly influenced by the Coulomb behavior in the low-energy region.
As the phase shift for the AV18 potential decreases rapidly with decreasing energy, reaching values of the order of $10^{-10}$ at 1~keV, the nuclear contribution to the scattering wave function becomes negligible compared with the Coulomb contribution. Consequently, the cross section is determined by the overlap between the deuteron bound-state wave function and the Coulomb wave function of the $pp$ scattering state. Since the fusion process is driven by the nuclear interaction at short distances, a cross section evaluated from an almost purely Coulombic scattering state may not adequately represent the low-energy fusion process. The cross sections obtained using the AV18 potential are identical to those reported in Ref.\cite{carlson1991weak}.

In contrast, the present calculations exhibit a different behavior. The phase shift decreases slowly with decreasing energy and remains finite, with a value of approximately $1.2^o$ even at an energy of 2~keV. Unlike the AV18 case, the phase shift does not approach zero in the low-energy region. Consequently, the scattering wave function continues to retain a significant nuclear contribution, leading to a continuous increase in the calculated radiative capture cross section as the energy decreases towards zero. From a physical viewpoint, as the relative energy decreases, the pp system approaches the formation of the bound deuteron state. Since the deuteron is a bound state with an energy of approximately $-2.2246~MeV$, it is physically reasonable to expect the radiative capture cross section to continue increasing rather than decreasing. As a consequence, the astrophysical (S)-factor also retains its energy dependence and does not approach a constant value down to 0.1~keV, as evident from the results presented in Table~\ref{tab1k}. Therefore, the Gamow peak is determined by the combined energy dependence of the astrophysical $S$-factor, the Gamow factor, and the Maxwell-Boltzmann distribution, rather than by the Gamow factor and the Maxwell-Boltzmann distribution alone. Alternatively, the Gamow peak can be obtained directly from the calculated radiative capture cross sections by numerically evaluating the overlap integral at low energies.

\begin{figure}[H]
\centering
\includegraphics[width=0.5\linewidth]{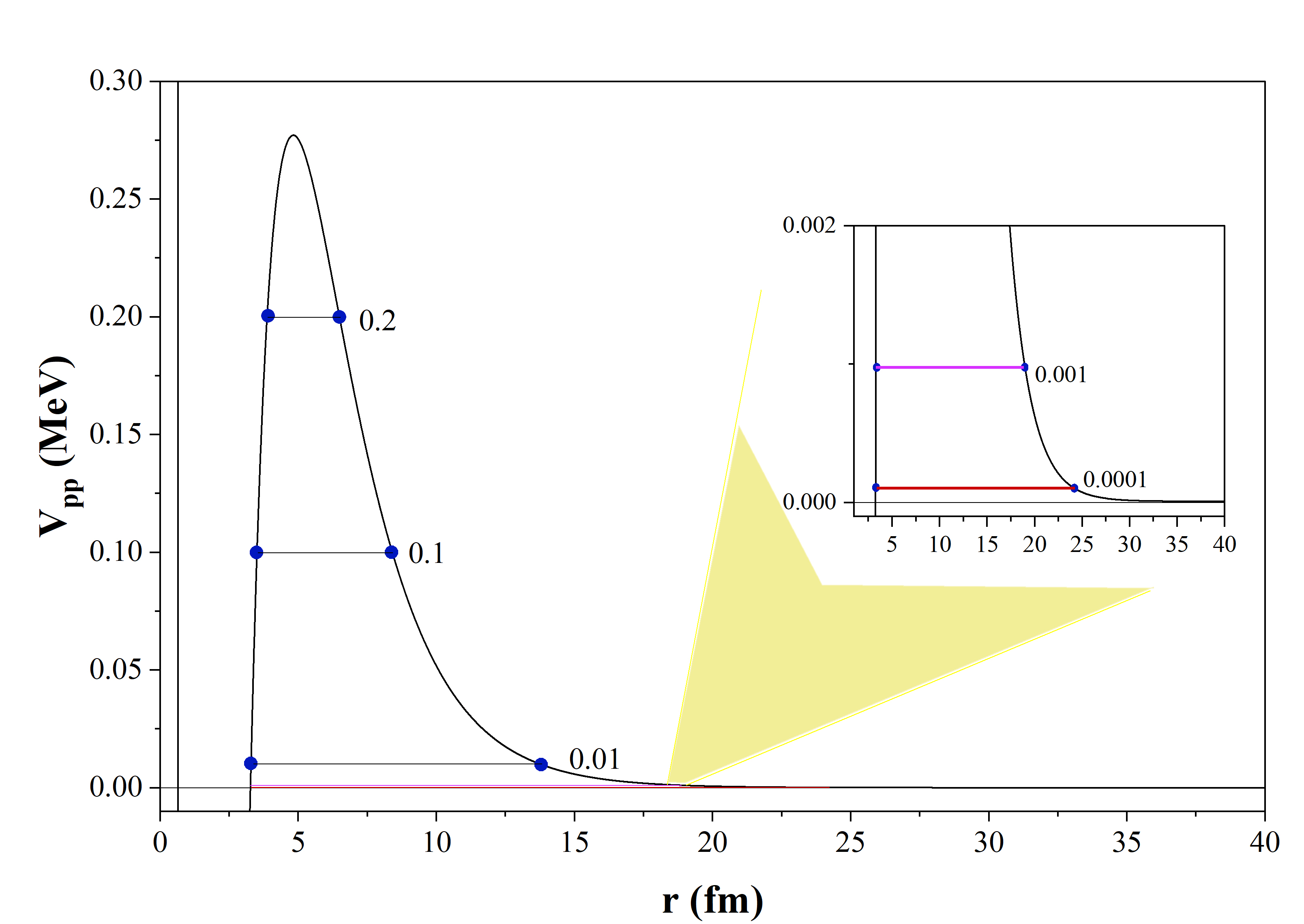}
\caption{Effective proton-proton interaction potential obtained within the Reference Potential Approach together with the corresponding classical turning points for different incident energies. The inset shows the enlarged low-energy region of the effective interaction potential, illustrating the determination of the tunneling region at extremely small energies.}
\label{cb}
\end{figure}

The effective proton–proton interaction potential associated with one of the optimized parameter sets obtained from the reference potential approach, together with the corresponding barrier penetration for different incident energies, is illustrated in Fig. \ref{cb}. It can be seen from the figure that the effective interaction potential gradually approaches zero at distance of approximately 32 fm, beyond which the interaction becomes negligible. Therefore, the classical turning points can be readily determined for incident energies ranging from 0.2 MeV to 0.0001 MeV. The WKB penetration probabilities corresponding to these energies are then used to compute the astrophysical S-factor empirically and subsequently extrapolated to determine the S-factor at zero energy, S(0). The horizontal lines represent the incident energies, while the corresponding intersection points with the effective interaction potential define the classical turning points used in the evaluation of the WKB tunneling integral. Using these turning points, action integral is evaluated numerically for each incident energy in order to determine the corresponding tunneling probability through the effective barrier.
The inset shows the enlarged low-energy region of the effective interaction potential. It demonstrates that even at extremely small energies such as E = 0.0001 MeV, the classical turning points and the corresponding WKB action integral can still be determined directly from the effective interaction potential without introducing an explicit bare Coulomb approximation. This effective interaction potential, obtained through the Reference Potential Approach, naturally accommodates the screening effects within the proton–proton interaction and therefore provides a consistent description of the low-energy tunneling process. \\

\textbf{Radiative Capture Calculations in the 0.5 - 0.1 keV Energy Range:}\\
The accuracy of the extrapolated zero-energy astrophysical S-factor, S(0), can be significantly improved by considering incident energies as low as possible. The use of sufficiently low-energy data points leads to a more reliable extrapolation procedure by minimizing the uncertainties arising from the energy dependence of the S-factor. Consequently, extrapolation from the lowest possible energies provides a better prediction of S(0), which is essential for accurately determining thermonuclear reaction rates in stellar environments and for understanding nucleosynthesis processes in astrophysical systems. Obtained values of S-factor along with the overlap integral and cross-section for the low-energy range up to 0.0001 MeV is given in Table~\ref{tab:sfactorm}.

\begin{table}[h]
\centering
\caption{Astrophysical (S)-factor values and corresponding uncertainties for low-energy proton–proton fusion, calculated using Eq.~\ref{sf}, based on interaction potentials obtained through the Reference Potential Approach.}
\label{tab:sfactorm}
\begin{tabular}{llllll}
\hline
E(lab) & {$\delta(degrees)$} &
{$e^{2I}$} &
{Overlap} &
{$\sigma(barn)$} &
{S(MeV-barn)} \\
\hline
0.0005 & 0.6086 $\pm$ 0.0375 & 3.3031$\pm$ 0.1582 & 0.0786$\pm$ 0.0013 & $3.2678 \pm 0.1136 \times 10^{-23}$  & $2.6948 \pm 0.0655 \times 10^{-26}$ \\
0.0004 & 0.5443 $\pm$ 0.0336 & 3.3170 $\pm$ 0.1596 & 0.0704$\pm$ 0.0012 & $3.6690 \pm 0.1226\times 10^{-23}$ & $2.4309 \pm 0.0582 \times 10^{-26}$ \\
0.0003 & 0.4713 $\pm$ 0.0291 & 3.3331 $\pm$ 0.1612 & 0.0602 $\pm$ 0.0013 & $4.1226 \pm 0.1818\times 10^{-23}$ & $ 2.0580 \pm 0.0643 \times 10^{-26} $ \\
0.0002 & 0.3848 $\pm$ 0.0237 & 3.3524 $\pm$ 0.1631 & 0.0486 $\pm$ 0.0013 & $4.9393 \pm 0.2706 \times 10^{-23}$ & $1.6531 \pm 0.0679 \times 10^{-26} $ \\

\hline
\end{tabular}
\end{table}

As evident from Table~\ref{tab:sfactorm}, the overlap integral decreases monotonically, whereas the WKB tunnelling integral I increases due to the increasing width of the tunnelling barrier. The calculated overlap integral is used to evaluate the nuclear matrix element through Eqn.~\ref{matrix}, which is subsequently employed in the calculation of the fusion cross-section using Eqn.~\ref{pcs}. Despite the variations in the calculated quantities, the astrophysical S-factor exhibits a smooth and monotonic decrease. Although the radiative capture cross section continues to increase at low energies, the astrophysical S(E) factor decreases monotonically with decreasing energy after the dominant Coulomb barrier penetration factor has been removed. The uncertainties quoted in the table correspond to the total error, which includes contribution of 11 potentials from the LOOCV analysis 

\begin{figure}[h]
    \centering
    \includegraphics[width=0.5\linewidth]{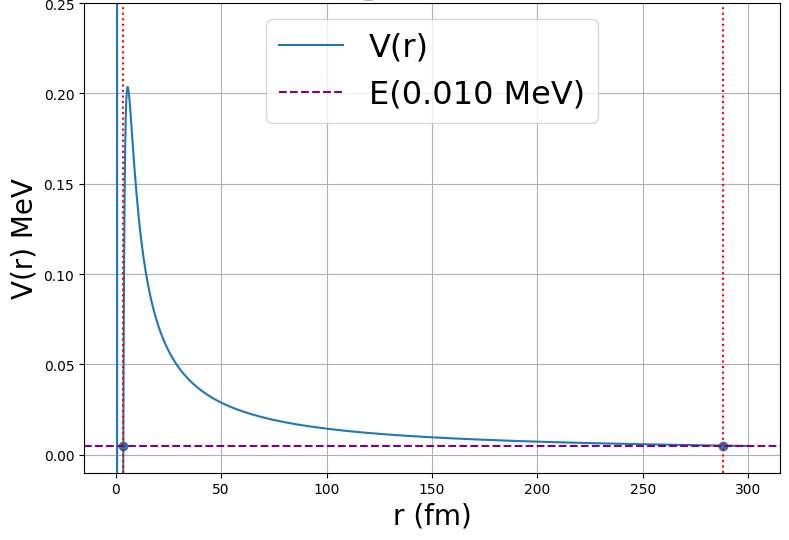}
    \caption{AV18 interaction potential as a function of radial distance for proton–proton fusion. The horizontal dashed line
corresponds to the center-of-mass energy E = 0.010 MeV, while the intersection points with the potential determine the
classical turning points within the WKB framework.}
    \label{Av18plot}
\end{figure}

For the AV18 potential, however, the numerical procedure is unable to locate the second classical turning point for energies below 0.01 MeV, even when the radial range is extended up to r = 300 fm. This behavior arises from the slowly decaying Coulomb tail, which shifts the outer turning point to very large distances at low incident energies. At E = 0.001 MeV, the outer turning point lies beyond the numerical integration limit of 300 fm, and therefore the WKB penetration integral cannot be evaluated reliably, as shown in Fig~\ref{Av18plot}. As a result, the Gamow exponent, overlap integral, and astrophysical S-factor cannot be determined at this energy within the present computational framework. This behavior highlights the numerical difficulty encountered in low-energy proton-proton fusion calculations when the Coulomb interaction is treated without incorporating
an explicit screening effect.

\begin{figure}
    \centering
    \includegraphics[width=0.5\linewidth]{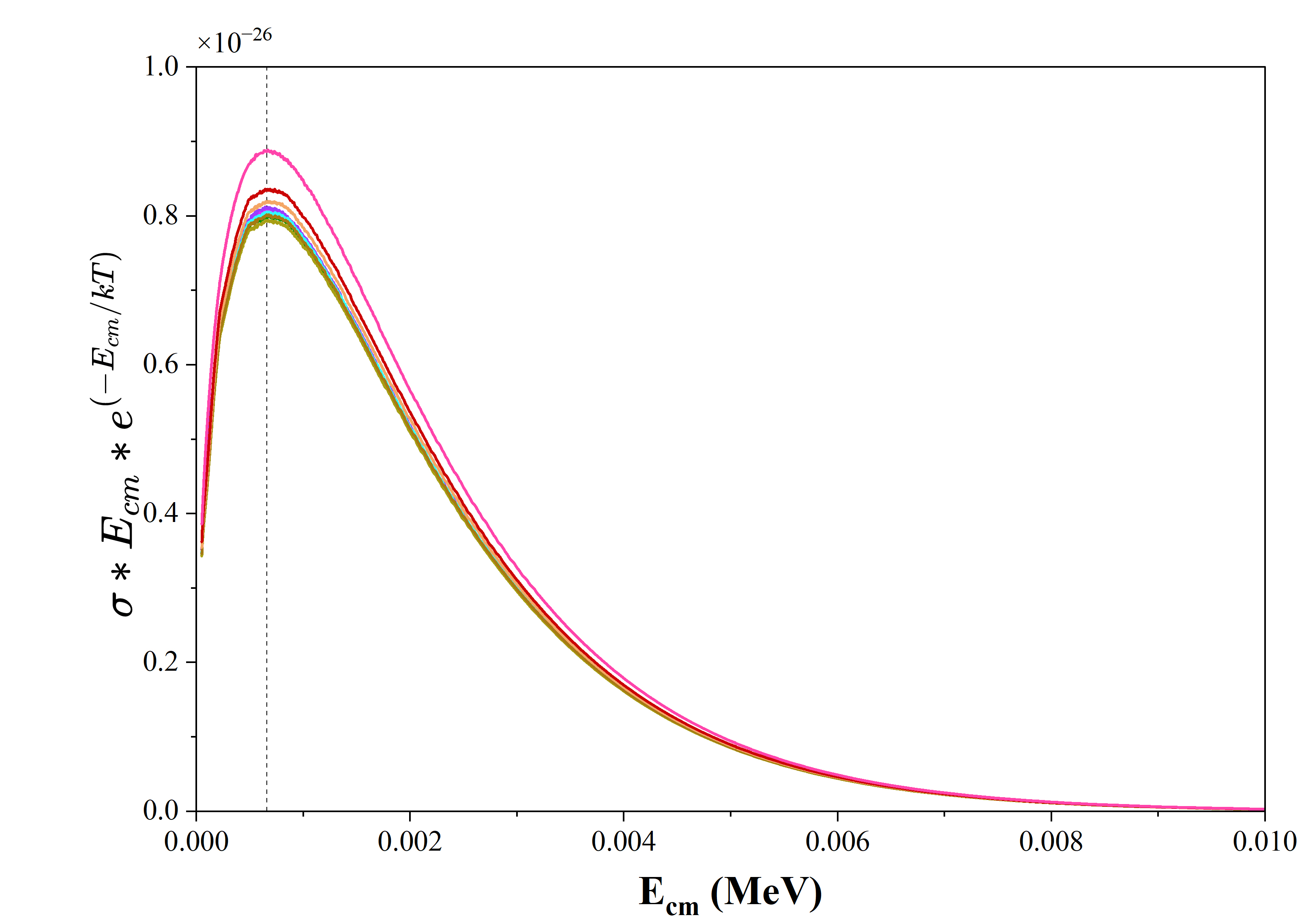}
    \caption{Gamow peak calculated using the cross section obtained from our overlap integral calculations.}
    \label{gamow}
\end{figure}

\textbf{Gamow energy peak:}\\
We have calculated the Gamow peak energy using the fundamental WKB action integral where V(r) is the total effective interaction potential obtained from reference potential approach. In Eqn.~\ref{ge}, using the Boltzmann constant ($k_B = 8.6173 \times 10^{-11} \, MeV\, K^{-1}$), and a temperature of ($T = 1.57 \times 10^7\, K$), the Gamow peak energy is calculated to be $0.67\,\pm\,0.02$ keV. However, literature reports a Gamow peak energy of $\approx$ 5 - 6 keV \cite{rolfs1988cauldrons}, obtained by expressing the WKB tunnelling integral in terms of the Sommerfeld parameter, $2\pi\eta$, derived from the bare Coulomb potential. This indicates that the choice of potential and the treatment of the tunnelling probability can significantly influence the estimated Gamow peak energy. If we intend to obtain the same value of the Gamow peak energy as reported in the literature, the temperature should be increased by more than one order of magnitude. This suggests that either the calculation of the Gamow peak energy needs to be reconsidered or that the temperature within the Sun may be higher than currently assumed. In the present calculations, both approaches (using either S-factor or cross-section) yield the same peak energy of approximately 0.6~keV, as shown in Fig.~\ref{gamow}.\\

\textbf{Neural Network-Based Extrapolation of the Astrophysical S-Factor:}\\
In the present analysis, instead of employing conventional polynomial extrapolation techniques, a supervised neural network \cite{Haykin1994,Goldberg1989} approach is used for the zero-energy extrapolation of the astrophysical \(S\)-factor . The neural network is trained using 1000 calculated \(S(E)\) values obtained over the energy interval from \(0.2\) MeV to \(0.0001\) MeV. Through the training procedure, the network learns the intrinsic nonlinear behavior and underlying energy dependence of the calculated fusion data directly from the physical input rather than assuming any predefined polynomial structure.
Unlike polynomial fitting methods, the neural network does not constrain the extrapolated behavior to a fixed analytical functional form. Instead, it constructs the extrapolation through adaptive learning of the correlations present in the calculated low-energy data. This allows the neural network to capture the subtle variations arising from the effective interaction potential, overlap integral, deuteron binding energy, and WKB tunneling probability in a more flexible and physically consistent manner.
The supervised learning framework therefore provides a more reliable extrapolation toward the zero-energy region, where the astrophysical \(S\)-factor plays a crucial role in solar-model calculations and stellar nucleosynthesis studies. By learning directly from the calculated low-energy fusion data, the neural network minimizes the uncertainties associated with arbitrary polynomial-order selection and avoids the artificial distortions that can arise in conventional fitting procedures.

\begin{figure}[h]
    \centering
    \includegraphics[width=0.4\linewidth]{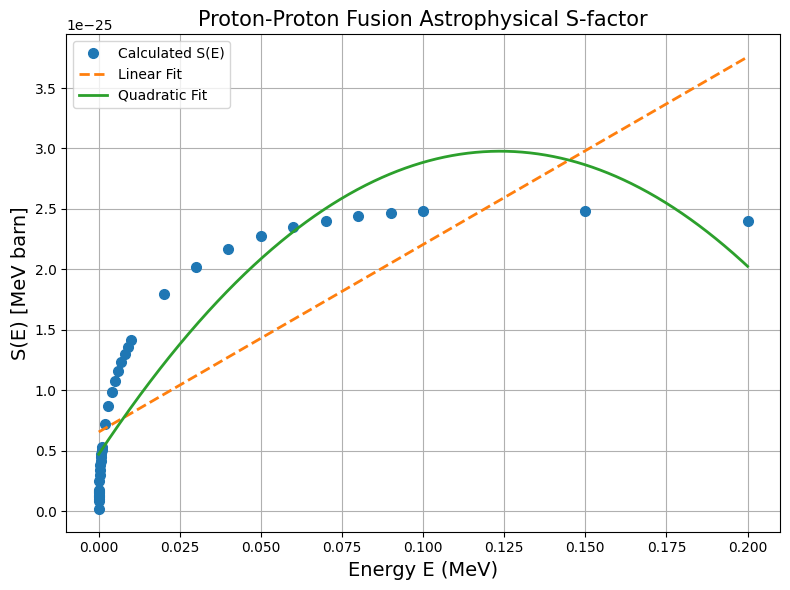}
    \includegraphics[width=0.43\linewidth]{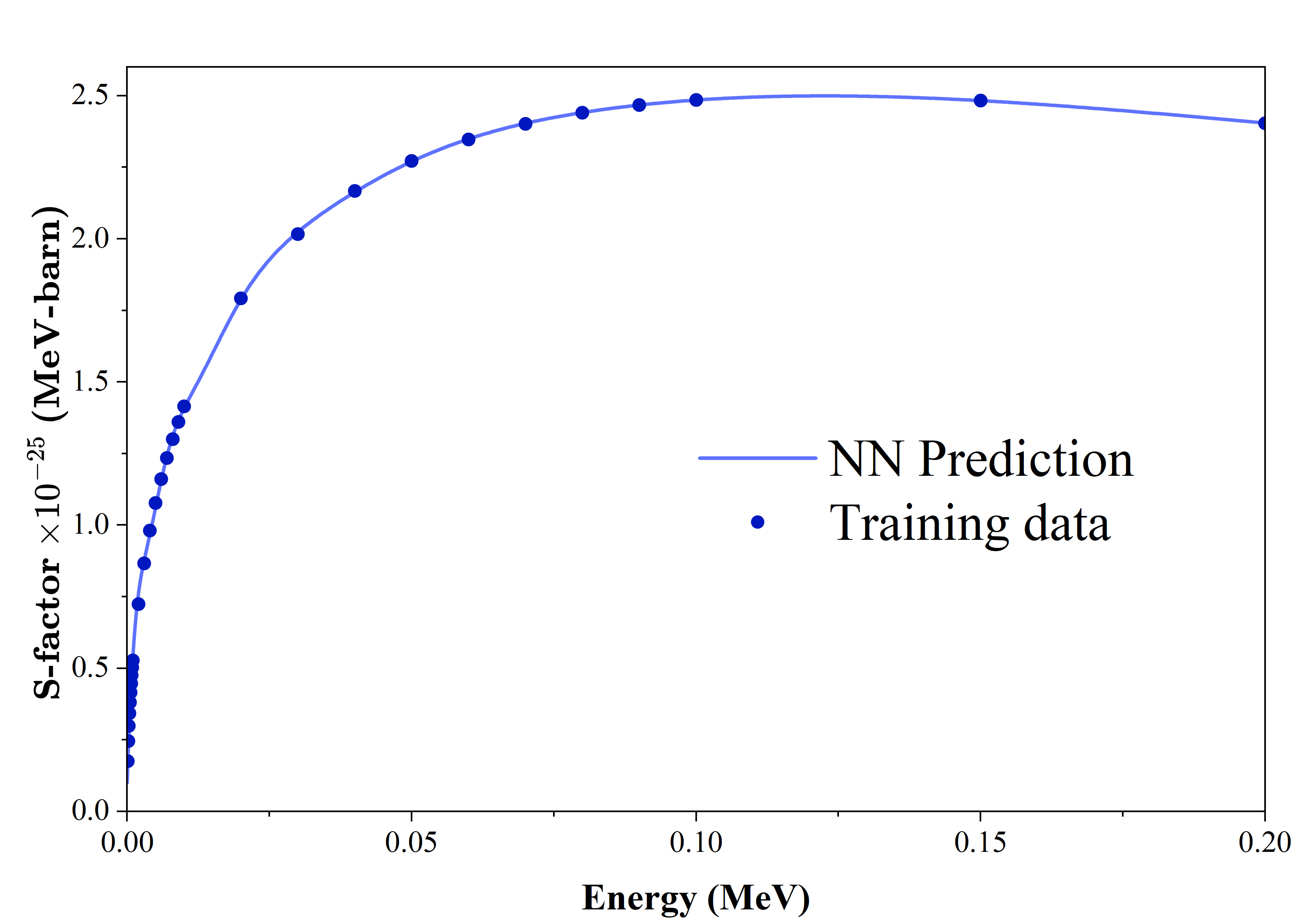}
    \caption{Comparison of the linear and quadratic extrapolation methods for the astrophysical S-factor using calculated data over the energy range $0.0001\leq E \leq0.2$ MeV (left), together with the neural network-based extrapolated value of S(0)(right)}
    \label{extrap}
\end{figure}

The extrapolations obtained using linear, quadratic, and neural network approaches are shown in Fig.~\ref{extrap}. It is evident from the figure that neither the linear nor the quadratic fit capture the data points appropriately over the considered energy range. Consequently, an extrapolation based solely on the quadratic fit may lead to an unreliable estimate of the zero-energy astrophysical S-factor, (S(0)). In contrast, the neural network accurately captures the underlying nonlinear dependence of (S(E)) on the incident energy without assuming any predetermined functional form. As a result, the extrapolated curve remains continuous and smooth, reducing the artificial fluctuations that may arise from low-order polynomial fits. This flexibility allows it to preserve the physical trend of the calculated S-factor more accurately during extrapolation, thereby providing a more reliable and robust estimate of S(0)

Using this neural-network-based extrapolation together with the eleven optimized LOOCV-generated interaction potentials, the average value of the astrophysical \(S\)-factor for the proton--proton fusion reaction is obtained as

\[
S(0)=(1.2607 \pm 0.0429)\times10^{-26}\ \mathrm{MeV\,b}.
\]

Here, the error represents the uncertainty calculated from the LOOCV sensitivity analysis. The value obtained in the present work is lower than the corresponding value reported in the literature. This discrepancy can be attributed to the difference between the Gamow factor calculated using the Sommerfeld parameter and those obtained from the WKB integral. The effect of this difference is clearly reflected in the calculated astrophysical S-factor value also. Since the S-factor is highly sensitive to the barrier penetration probability, even a small variation in the evaluation of the Gamow factor can result in significant changes in the extracted S-factor, particularly in the low-energy region relevant to proton-proton fusion.

The comparatively smaller value of the S-factor obtained in the present analysis indicates that several previously reported values available in the literature may be slightly overestimated due to the use of simplified Coulomb approximations and asymptotic treatments of the tunneling probability. The present results provides a second evidence that the actual temperature of the solar core could be higher than the values assumed in earlier studies. Because nuclear reaction rates and the astrophysical S-factor depend strongly on temperature, even a slight increase in the core temperature can substantially affect the predicted fusion rates and, consequently, the energy production within the Sun. 


One limitation of the inverse approach is the non-uniqueness of the reconstructed potential, since different phase-equivalent potentials may reproduce the same scattering phase shifts. Although the cutoff radius is determined by the inversion procedure, its value may vary among different reconstructed potentials, leading to slight differences in the resulting potential parameters. Furthermore, the present calculation includes only the one-body weak current. The contribution of the two-body axial current, which is expected to slightly modify the nuclear matrix element, has not been included in the present analysis and remains a subject for future investigation. The present work offers two major improvements over the conventional approach. First, the tunnelling probability is evaluated using the effective interaction potential, thereby ensuring consistency with the potential employed in the wave-function calculation instead of relying on the bare Coulomb approximation. Second, the extrapolation of the astrophysical S-factor to zero energy is performed using a neural network rather than a quadratic fit, providing a more flexible representation of the energy dependence of S(E).

The present result therefore suggests that a fully consistent treatment of the proton–proton interaction, overlap integral, and low-energy tunneling dynamics plays an important role in obtaining reliable estimates of the solar proton–proton fusion rate.

\section*{Conclusion}

     The present work provides a unified and self-consistent framework for the determination of the astrophysical S-factor for proton-proton fusion in the extremely low-energy region relevant to stellar interiors.

\begin{itemize}

\item Our methodology of reference potential approach has been successful in contributing an interaction potential for pp scattering that is phenomenologically inclusive of nuclear, Coulomb, and screening. This paved the way to determine the actual Gamow penetration factor, using WKB integral for the first time, thus avoiding the bare Coulomb approximation that tends to completely negate the nature of the interaction potential.

\item The implication of the obtained Gamow penetration factor directly impacts the Gamow peak energy and thus reflects an increase of more than one order of magnitude in the solar core temperature

\item Another important consequence from our quickly decaying Coulomb barrier is the capability to determine the overlap integral and, from it, the capture cross-section to extremely low energies. This enables the determination of S-factor for energies as low as 0.0001 MeV with greater accuracy, even though extrapolation error due to phase shift remains.

\item The conventional polynomial fitting procedures have been replaced by a supervised neural network to learn the inherent non-linear relationship between E and S(E) thus obtaining S(0) as $0.1261\times 10^{-25}$ which is observed to be less than one order of magnitude than the currently accepted values.

\end{itemize}
Our methodology for calculating S-factor using interaction potentials obtained through RPA can be extended to other astrophysical reactions such as deuteron-deuteron capture, which will have implications with regard to nuclear fusion reactors.

\bibliographystyle{unsrt}

\bibliography{references}

\end{document}